\documentclass[onecolumn,11pt,a4paper,aps]{revtex4-2}

\newcommand{\eq}[1]{\begin{equation}#1\end{equation}}
\newcommand{\dd}{\mathrm{d}}
\newcommand{\ex}{\mathrm{e}}

\newcommand{\Tr}{\mathrm{Tr}}

\def \be {\begin{equation}} 
\def \ee {\end{equation}}

\def \la {\langle} 
\def \ra {\rangle}  

\usepackage{amsmath}
\usepackage{amsfonts}
\usepackage{amssymb}
\usepackage{braket}
\usepackage{bbold}
\usepackage{comment}
\usepackage{graphics}
\usepackage{graphicx}
\usepackage{leftidx}
\usepackage{psfrag}
\usepackage{color}
\usepackage{colordvi}
\usepackage{bbm}
\usepackage{hyperref}

\begin{document}

\title{Zero-mode entanglement across a conformal defect}

\author{Luca Capizzi$^1$ and Viktor Eisler$^2$}

\affiliation{
$^1$SISSA and INFN Sezione di Trieste, via Bonomea 265, I-34136 Trieste, Italy\\
$^2$Institute of Theoretical and Computational Physics, Graz University of Technology,
Petersgasse 16, A-8010 Graz, Austria
}

\begin{abstract}

We consider a free-fermion chain with a conformal defect that features an extended zero mode,
and study the entanglement properties in its mixed ground state. The zero-mode induced degeneracy modifies the density of states in the single-particle entanglement spectrum, which can be calculated via the full counting statistics. For a homogeneous chain, the resulting change in the R\'enyi entropy is derived analytically for arbitrary subsystem ratios in the thermodynamic limit. For a conformal defect located in the center, analogous results can be obtained for the half-chain entanglement. In particular, we observe parity effects for half-chains with even/odd sites, which do not decay with size.
\end{abstract}

\maketitle

\section{Introduction}

Entanglement in quantum many-body systems has developed  into a major research field, receiving attention from many communities ranging from high-energy to condensed-matter physics \cite{afov-08,ecp-10,ccd-09,Laflorencie-16}. Among the various aspects, numerous studies have been devoted to the characterization of entanglement in one-dimensional critical systems, which can be described by 1+1D conformal field theories (CFT) \cite{cc-09}. In particular, the entanglement entropy of a segment was found to scale logarithmically with its size, with a prefactor that is proportional to the central charge of the CFT \cite{hlw-94,cc-04}. The logarithmic entropy growth was observed also in critical spin chains \cite{vlrk-03}, and the corresponding
violation of the area law \cite{ecp-10} was recognized as a distinct feature of criticality.

While the dominant contribution to the entropy turns out to be universal, this is, in general, not the case for the subleading term. Nevertheless, there are examples where this term contains some universal piece that is finite and independent of the UV cutoff, such as the case of 1D systems with a boundary \cite{ALS09}. Indeed, this contribution is entirely determined by the boundary CFT, and is related to the boundary entropy of Affleck and Ludwig \cite{AL91}. Such boundary entropies have been studied extensively both within CFT \cite{cc-04,txas-13,eir-22,eir-23,top-16,mt-20} as well as numerically for various critical spin chains in the presence of boundary fields \cite{LSCA06,XR20,RS22b}.

Some particular choice of boundary conditions can also lead to the emergence of zero modes. These are extended excitations which have exactly zero energy, and thus yield a degeneracy of the spectrum. For a single zero mode, the proper ground state of the system is then a mixture of two pure states, with the zero mode either empty or occupied. Interestingly, the entropy of the mixed ground state differs from the pure state one by a nontrivial function of the subsystem ratio, which was calculated analytically for free Dirac or Majorana fermions \cite{hn-13,KVW17,KVW18}. The result has later been verified for a quantum Ising chain with a topological defect \cite{RS22,rpr-22}, which describes a special boundary condition that produces a zero mode \cite{amf-16}. One should stress that this zero mode is an extended excitation, and should not be confused with the ones that are localized at the boundaries, as found in gapped phases of various quantum chains \cite{Kitaev01,Fendley12,Fendley16}.

The topological defect corresponds to perfect transmission and thus reproduces the zero-mode entropy found for a periodic chain \cite{KVW17}. Here we address the question, how the zero-mode contribution is altered for a defect with imperfect transmission. In fact, the presence of a defect at the subsystem boundary in free-particle chains is known to modify the logarithmic scaling of the entropy, leading to a prefactor (also dubbed as effective central charge) which depends on the transmission properties of the defect. This was first investigated numerically in free-fermion and transverse Ising chains \cite{peschel-05,isl-09,eg-10}, and the analytic expression of the effective central charge was found in \cite{ep-10,pe-12}. On the CFT side, the entanglement across conformal defects was studied for the free boson \cite{ss-08} and the Ising model \cite{bb-15}, perfectly matching the lattice results. Generalizations for CFT junctions with multiple wires were considered in \cite{gm-17,cmc-22,cmc-22a}.

In this paper we consider a free-fermion chain with a conformal defect which supports an exact zero mode,
and study the corresponding finite contribution to the half-chain entropy in its mixed ground state.
The conformal defect on the lattice mimics the scale-invariant properties of a conformal interface in CFT \cite{bbdo-02}, and was first studied in \cite{EP12}. Most importantly, it allows one to establish an exact relation between the half-chain entanglement spectrum of the defect as well as that of the homogeneous chain. We use this relation to derive an analytical prediction for the zero-mode entropy, thus generalizing the studies of Ref. \cite{RS22} to a chain with imperfect transmission. In particular, our result shows parity effects in terms of the half-chain length, which vanish only when the defect is completely transmissive. Our analytical predictions, derived in the thermodynamic limit, are in perfect agreement with the numerical results.

We organize our manuscript as follows. In Sec. \ref{sec:model} we introduce the model and the methods employed. In Sec. \ref{sec:hom} we characterize the zero-mode entropy in the homogeneous chain for arbitrary subsystem ratios, reproducing the results of \cite{KVW17} in an alternative way. In Sec. \ref{sec:Defect} we compute the zero-mode entropy of the half-chain in the presence of a conformal defect. We summarize and discuss our results in Sec. \ref{sec:discussion}, leaving some technical details of the calculations in three appendices.

\section{Model and methods\label{sec:model}}

We consider hopping chains described by the Hamiltonian
\eq{
\hat H = \sum_{m,n} H_{m,n} \, c_m^\dag c^{\phantom{\dagger}}_n ,
\label{H}}
where $c^\dag_m$ and $c_m$ are fermionic creation/annihilation operators satisfying
anticommutation relations $\{ c^\dag_m, c_n \} = \delta_{m,n}$. We shall focus on models
with only nearest-neighbour hopping and local chemical potentials.
The Hamiltonian \eqref{H} can be diagonalized by finding the eigenvalue decomposition
of the real and symmetric hopping matrix
\eq{
H_{m,n} = \sum_{k} \omega_k \, \phi_k(m)\phi_k(n) \, ,
\label{Hmn}}
with single-particle spectrum $\omega_k$ and corresponding eigenvectors $\phi_k$.
In general, the ground state of the chain is a Fermi sea with occupied modes $\omega_k<0$ for $k<k_F$.
However, the situation changes if the system supports a zero mode, i.e. one has an
eigenvalue $\omega_{k_F}=0$ in the spectrum. In this case, the mode occupation should
be determined as a proper zero-temperature limit of the Fermi function
\eq{
n_k = \lim_{\beta\to\infty} \frac{1}{e^{\beta \omega_k}+1}=
\begin{cases}
1 \quad \omega_k < 0 \, ,\\
\frac{1}{2} \quad \omega_k = 0 \, , \\
0 \quad \omega_k > 0 \, . \\
\end{cases}
\label{nk}}
In other words, due to the zero mode and the resulting degeneracy of the spectrum,
the ground state $\hat\rho$ becomes mixed, given by an equal superposition
\eq{
\hat\rho = \frac{1}{2}(\hat\rho_0 + \hat\rho_1)
\label{rho}}
of two pure Fermi-sea ground states $\hat\rho_0=\ket{0}\bra{0}$ and $\hat\rho_1=\ket{1}\bra{1}$, where
\eq{
\ket{0} =\prod_{k<k_F} c_k^\dag \ket{\emptyset}, \qquad
\ket{1} = \prod_{k \le k_F} c_k^\dag \ket{\emptyset},
%\ket{1} = c^\dag_{k_F} \ket{0}
\label{01}}
denote the states with the zero mode being empty or occupied, respectively. 
Here the empty state with no particles is denoted by $\ket{\emptyset}$, and $c^\dag_k$
are the creation operators in the diagonal basis of $\hat H$.

Our goal is to study the contribution of the zero mode to the entanglement between a
subsystem $A$ and its complement $B$. This is encoded in the reduced density
matrix $\hat\rho_A=\Tr_B \, \hat\rho$, which can be written in the form \cite{Peschel03,PE09}
\eq{
\hat\rho_A = \frac{1}{Z}
\exp \Big(-\sum_\kappa \varepsilon_\kappa \, f^\dagger_\kappa f^{\phantom{\dagger}}_\kappa \Big) .
\label{rhoA}}
Here $\varepsilon_\kappa$ is the single-particle entanglement spectrum, which can be related via
\eq{
\varepsilon_\kappa = \ln \frac{1-\zeta_\kappa}{\zeta_\kappa} \, , \qquad
\zeta_\kappa = \frac{1}{\ex^{\varepsilon_\kappa}+1}
\label{epszeta}}
to the eigenvalues $\zeta_\kappa$ of the reduced correlation matrix $C_A$ with elements
\eq{
C_{mn} = \braket{c^\dag_m c_n} = \sum_{k} n_k \, \phi_k(m)\phi_k(n) \, ,
\qquad m,n \in A \, .
\label{CA}}
Note that we use the symbol $\kappa$ to differentiate the modes of the entanglement Hamiltonian
in \eqref{rhoA} from those of the physical Hamiltonian in \eqref{Hmn}.

With the spectrum $\varepsilon_\kappa$ at hand, the entanglement entropy $S=-\Tr (\hat\rho_A \ln \hat\rho_A)$ 
is obtained as
\eq{
S = \sum_\kappa s(\varepsilon_\kappa) \, , \qquad
s(\varepsilon) = \frac{\varepsilon}{\ex^{\varepsilon}+1} +
\ln (1+\ex^{-\varepsilon}) \, .
\label{seps}}
In the thermodynamic limit of a very large subsystem, the spectrum becomes densely spaced and
the sum can be replaced by an integral
\eq{
S \to \int \dd \varepsilon \, \rho(\varepsilon) \, s(\varepsilon) \, , \qquad
\rho(\varepsilon) = \frac{\dd \kappa}{\dd \varepsilon} \, ,
\label{rhoeps}}
where $\rho(\varepsilon)$ is the density of states. In order to obtain analytical results for $\rho(\varepsilon)$
in the thermodynamic limit, it is useful to introduce the resolvent
\eq{
R(z) = \Tr \left( \frac{1}{z-C_A}-\frac{1}{z}\right).
\label{R}}
Using the Sokhotski-Plemelj formula of complex analysis, the spectral density of $C_A$
is related to the resolvent as
\eq{
\rho(\zeta) \equiv \Tr \left[ \delta(\zeta-C_A)-\delta(\zeta) \right] =
\frac{1}{2\pi} \lim_{\epsilon \to 0^+} \text{Im}
\left[ R(\zeta-i\epsilon) -  R(\zeta+i\epsilon) \right] .
\label{rhoz}}
Note that, for later convenience, a $1/z$ term has been subtracted in the definition
\eqref{R} of the resolvent, which leads to an additional $\delta(\zeta)$ contribution
in the spectral density \eqref{rhoz}. This additional term, however, does not play a role since
we will be interested in the difference of spectral densities.

The final step is to relate the resolvent to the full counting statistics (FCS),
which is just the probability distribution of the particle number
$\hat N_A=\sum_{n \in A} c_n^\dag c^{\phantom{\dagger}}_n$ within the subsystem $A$.
The cumulant generating function of the FCS reads
\eq{
\chi(\alpha) = \ln \Tr( \hat{\rho}_A \ex^{i\alpha \hat N_A}) = \Tr \ln  \left[1-(1-e^{i\alpha})C_A\right].
\label{chi}}
Introducing the variable
\eq{
z= \frac{1}{1-e^{i\alpha}},
\label{z}}
and considering the FCS as a function $\chi(z)$, one can immediately
see that the resolvent follows as
\eq{
R(z) = \frac{d}{dz} \chi(z) \, .
\label{Rchi}}
Hence, we have directly related the spectral density of the reduced correlation matrix
to the FCS, which has been extensively studied in the ground states of critical 1D systems \cite{kl-09,Songetal11,AIQ11,IAC13,SI13,cmc-23}.
Obtaining the density of the entanglement spectrum is then a simple change of
variables
\eq{
\rho(\varepsilon) = \left|\frac{\dd \zeta}{\dd \varepsilon}\right| \rho(\zeta) \, ,
\label{rhoepszeta}}
using the relations \eqref{epszeta}.

In the following section we study the zero-mode induced variation of the density of states in a homogeneous chain, and the resulting change in the entropy.

\section{Zero mode in a homogeneous chain\label{sec:hom}}

Let us first consider an open chain of even length $2L$ with some local chemical potentials at
its boundaries, such that the nonvanishing entries of the hopping matrix are given by
\eq{
H_{m,m+1}=H_{m+1,m}=-1/2 \, , \qquad
H_{1,1}=H_{2L,2L}=1/2 \, .
\label{Hhom}}
It is easy to show that the eigenvalues and vectors are
\eq{
\omega_k = -\cos \Big(\frac{\pi k}{2L}\Big)  \, , \qquad
\phi_k(m) = \mathcal{N}_k \, \sin\left[ \frac{\pi k}{2L}(m-1/2)\right],
\label{phihom}}
where $k = 1, \dots, 2L$ and the normalization factor is given by $\mathcal{N}_k=1/\sqrt{L}$ for $k\ne2L$
as well as $\mathcal{N}_{2L}=1/\sqrt{2L}$. One has thus a single zero mode with $k=k_F=L$, and the
squared amplitudes of the corresponding eigenvector are constant $\phi^2_{k_F}(m)=1/(2L)$.
Note that the existence of the zero mode is due to the special choice of boundary conditions in \eqref{Hhom}.
Instead, for an open chain of length $2L$ without boundary potentials, the momenta are quantized as $\pi k/(2L+1)$,
and the zero mode is absent. It would only appear for odd chain sizes, however, for a better analogy with
the defect problem in the next section, we prefer to work with an even number of sites.

The correlation matrix elements can be calculated explicitly using the eigenvectors in \eqref{phihom}.
In particular, the state $\hat\rho_1$ has $L$ occupied modes and one obtains
\eq{
C_{1,mn}=
\frac{\sin \left[\frac{\pi(2L+1)}{4L}(m-n)\right]}{4L\sin\left[\frac{\pi}{4L}(m-n)\right]}
-\frac{\sin \left[\frac{\pi(2L+1)}{4L}(m+n-1)\right]}{4L\sin\left[\frac{\pi}{4L}(m+n-1)\right]} \, .
\label{C1hom}
}
For the mixed-state correlations \eqref{CA} with the half-filled zero mode in \eqref{nk} one has
\eq{
C_{mn} = C_{1,mn} - \frac{1}{2L}\sin\left[ \frac{\pi}{2}(m-1/2)\right]\sin\left[ \frac{\pi}{2}(n-1/2)\right],
\label{Chom}}
and the matrix elements $C_{0,mn}$ for the state $\hat\rho_0$ are very similar, with an extra factor two multiplying the second term.
We are interested in the change of the entanglement entropy $\delta S = S-S_0$, 
which gives the contribution of the zero mode in the mixed state $\hat\rho$ in \eqref{rho}
as compared to the pure state $\hat\rho_0$. The entropies are calculated for a subsystem
$A=\left[1,\ell\right]$ via the correlation matrices $C_A$ and $C_{0,A}$, respectively,
using the methods introduced in section \ref{sec:model}.

In order to obtain our analytical results, we consider a thermodynamic limit by fixing the
ratio $r=\ell/(2L)$ and sending $L\to \infty$. We focus on extracting the difference of the
density of states $\delta\rho(\varepsilon)=\rho(\varepsilon)-\rho_0(\varepsilon)$ in the
entanglement spectra. It turns out that the key object we need is the ratio of the FCS
calculated for the pure states $\hat\rho_0$ and $\hat\rho_1$, with the zero mode either
empty or occupied. This can be obtained via bosonization and CFT techniques,
as shown in appendix \ref{app:CFT}, yielding the simple result
\eq{
\frac{\Tr (\hat\rho_{1} e^{i\alpha \hat N_A})}{\Tr (\hat\rho_{0} e^{i\alpha \hat N_A})} =
%{\bra{1} e^{i\alpha \hat N_A}\ket{1}} =
\frac{\det (1-(1-e^{i\alpha}) C_{1,A})}{\det (1- (1-e^{i\alpha}) C_{0,A})} = 
\ex^{i\alpha r} .
%\cos (\alpha r/2).
\label{detratio}}

The physical interpretation of \eqref{detratio} is that the inclusion of the zero mode simply shifts the mean particle number by $r$ in $A$, but does not affect any higher order cumulants in the limit $L\to\infty$. Note that it seems hard to derive \eqref{detratio} directly on the lattice for the geometry at hand.
Results for the FCS are available for an infinite chain \cite{AIQ11}, obtained via Fisher-Hartwig methods for Toeplitz determinants \cite{BT91}. One could generalize it to Toeplitz + Hankel matrices using the results of Ref. \cite{DIK11}, however, this would correspond to a semi-infinite chain with a segment $\ell \gg 1$ at the boundary. Although the asymptotics of the determinant in the FCS is not directly accessible when we fix the ratio $r$, one could check the cumulants directly. In particular, the particle number fluctuations are given by $\Tr\left[C_{\sigma,A}(1-C_{\sigma,A})\right]$ for $\sigma=0,1$, and we observe numerically that their difference vanishes slowly as $\ln L/L$ for $L\to\infty$, in an alternating fashion.

For the mixed state $\rho$, the change of the cumulant generating function is obtained via \eqref{detratio} as
%for the mixed state $\hat\rho$ one obtains a nontrivial change of the cumulant generating function
%
\eq{
\delta \chi(\alpha) = \chi(\alpha)-\chi_{0}(\alpha) =
\ln \left(\frac{1+\ex^{i\alpha r}}{2}\right).
%\ln \left[\ex^{- i\alpha r/2} \cos (\alpha r/2)\right].
\label{dchi}}
The difference of the resolvents $\delta R(z)=R(z)-R_0(z)$ can be obtained using \eqref{Rchi},
by first substituting the $z$ variable \eqref{z} and performing the derivative
\eq{
\delta R(z)= \frac{\dd}{\dd z} \delta \chi(z)=
-\frac{r}{z(1-z)}\frac{(1-z^{-1})^r}{1+(1-z^{-1})^r} \, .
% + \frac{r}{2}\left(\frac{1}{z}-\frac{1}{z-1}\right).
%\chi(z)-\chi_{1}(z)
\label{dR}}
The change of the spectral density then follows from the formula \eqref{rhoz}.
To carry out the limit, let us first note that the expression \eqref{dR} has a branch cut along $z\in[0,1]$,
and thus $\delta\rho(\zeta)$ is supported on this interval, as it should. To evaluate the jump
across the branch cut, we need the limit
\eq{
\lim_{\epsilon \to 0^+}(1-(\zeta \pm i\epsilon)^{-1})^r = e^{\pm i\pi r}(\zeta^{-1}-1)^r \, .
}
Plugging this into \eqref{dR} and \eqref{rhoz}, we arrive at
\eq{
\delta\rho(\zeta) = 
\frac{r}{\pi \zeta(1-\zeta)}\frac{\sin(\pi r)(\zeta^{-1}-1)^r}{1+2\cos(\pi r)(\zeta^{-1}-1)^r + (\zeta^{-1}-1)^{2r}} \, .
% +\frac{r}{2}\left( \delta(z)-\delta(z-1)\right),
\label{drhoz}}
Note that, apart from the branch cut, the resolvent \eqref{dR} has an extra pole at $z=0$. Indeed, from the $z\to 0$ behaviour $\delta R(z) \simeq - r/z$ one infers that there is an extra delta function $- r \delta(\zeta)$
appearing in the spectral density. However, since the entropy density vanishes at the spectral edge $\zeta=0$,
we will simply discard this contribution.

\begin{figure}[htb]
\center
\includegraphics[width=0.6\textwidth]{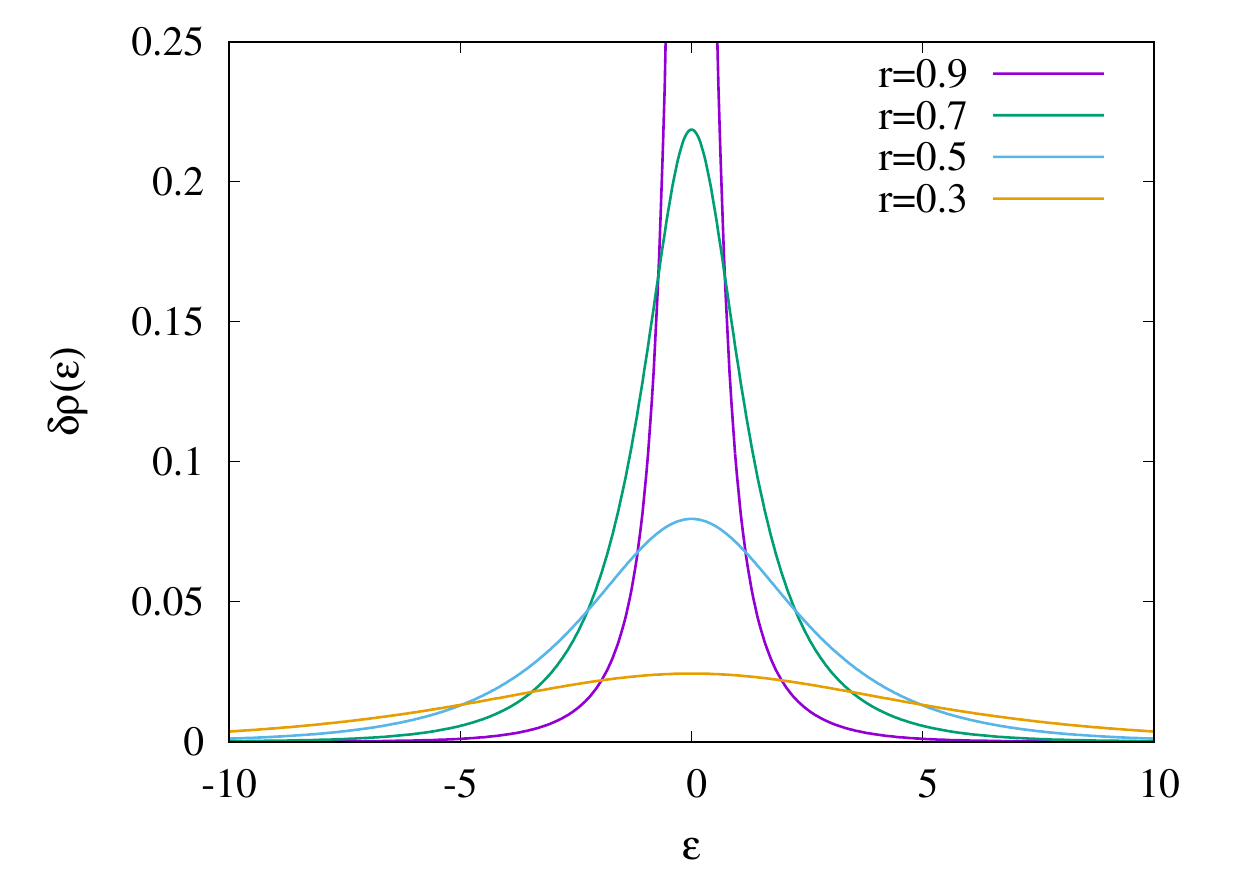}
\caption{Difference of the spectral density $\delta\rho(\varepsilon)$ for various values of $r$.
For $r \to 1$ the density converges to a delta function localized at $\varepsilon=0$.}
\label{fig:drhoeps}
\end{figure}

The change in the entanglement spectrum density is obtained via \eqref{rhoepszeta} by a change of variables
\eq{
\delta\rho(\varepsilon) = \frac{r}{2\pi}\frac{\sin(\pi r)}{\cos(\pi r)+\cosh(\varepsilon r)} \, .
\label{drhoeps}}
The spectral density $\delta\rho(\varepsilon)$ is shown in Fig. \ref{fig:drhoeps} for various ratios $r$.
One observes that the density becomes more and more peaked around $\varepsilon=0$ as one
increases the ratio towards $r \to 1$. Indeed, in this limit one can expand \eqref{drhoeps} to get
\eq{
\delta\rho(\varepsilon) \simeq \frac{1}{\pi}
\frac{\pi(1-r)}{\pi^2(1-r)^2 + \varepsilon^2} \underset{r \to 1}{\rightarrow}\delta(\varepsilon),
}
such that it precisely reproduces a delta function. In terms of the correlation matrix spectrum,
it corresponds to the appearance of an eigenvalue $\zeta=1/2$. Obviously, this is simply the half-filled
mode in \eqref{nk}, as in the limit $r \to 1$ of a full system one has $\zeta_\kappa=n_k$ for the eigenvalues of $C$.
One should also remark that the entropy difference $S-S_1$ measured from the state with an occupied zero mode
produces exactly the same results. Indeed, this simply corresponds to a change $r \to -r$ in \eqref{dchi},
but the final result \eqref{drhoeps} is manifestly symmetric under this transformation.
In our numerical calculations presented in the next subsections we actually used the convention
$\delta S = S-S_1$.

%%%%%%%%%%%%%%%%%%%%%%%%%%%%%
%%%%%%%%%%%%%%%%%%%%%%%%%%%%%

\subsection{Zero-mode entropy}

We can now apply the above results to calculate the zero-mode contribution to the entropy. Using the entropy density \eqref{seps} as well as the density of states \eqref{drhoeps}, one arrives at the integral
\eq{
\delta S = \int_{-\infty}^{\infty} \dd \varepsilon \, \delta\rho(\varepsilon) \, s(\varepsilon) \, .
\label{dShom}}
It is instructive to compare the above expression to the one derived in Ref. \cite{KVW17}
for chiral fermions in a ring geometry, which reads
\be
\delta S =  \pi r \int^\infty_0 dh \tanh(\pi h r)(\coth(\pi h)-1) \, .
\label{dSKVW}
\ee
Using the symmetry for $\varepsilon \to -\varepsilon$, the integrals \eqref{dShom} and \eqref{dSKVW} are defined on the same domain but with integrands that do not match.
Remarkably, however, a numerical evaluation of the integrals shows, that they reproduce the exact same function $\delta S(r)$. 

The origin of this mismatch can be understood as follows. The derivation in \cite{KVW17} is also based on the resolvent, and one can actually verify that the result for $\delta R(z)$ is exactly the same as ours in \eqref{dR}. However, the entropy is then extracted by making use of the formula
\eq{
\delta S= \int_{1}^{\infty} \dd z \, (1-z) \, [\delta R(z)-\delta R(1-z)] \, ,
\label{dSres}}
and a subsequent change of variables $h=\frac{1}{2\pi}\ln(\frac{z}{z-1})$ leads exactly to the result \eqref{dSKVW}. In other words, \eqref{dSres} uses the analytical regime of the resolvent to reproduce $\delta S$ by a mathematical trick. Indeed, performing the integration over $z$ before taking the trace in the definition of the resolvent $\delta R$, one obtains $\delta S = S-S_1$ with
\eq{
S = \Tr [ -C_A \ln C_A -(1-C_A) \ln (1-C_A)],
}
and similarly for $S_1$ using $C_{1,A}$. This is exactly the free-fermion formula for the entropies, and thus $\delta S$ is now obtained without ever referencing the spectral density $\delta \rho(\zeta)$. Note also the analogy between the change of variables $\zeta \to \varepsilon$ and $z \to h$.

A clear advantage of the representation \eqref{dShom} via the spectral density is that one can directly generalize it to evaluate the R\'enyi entropies
\eq{
S_n = \frac{1}{1-n} \ln \Tr(\hat\rho^n_A),
}
and the corresponding zero-mode contributions $\delta S_n = S_n-S_{1,n}$. Inserting the expression for the R\'enyi entropy density in terms of $\zeta$, one has
\be
\delta S_n = \frac{1}{1-n}\int^{1}_0 \dd \zeta \, \delta \rho(\zeta) \log\left[\zeta^n+(1-\zeta)^n\right],
\label{Sn}
\ee
which can be evaluated numerically for arbitrary R\'enyi index $n$. However, a considerable
simplification occurs for integer indices $n\geq 2$. Indeed, in such cases there is a well-known
relation between the R\'enyi entropy and the FCS \cite{CFH05}
\be
S_n = \frac{1}{1-n}\sum^{\frac{n-1}{2}}_{p = -\frac{n-1}{2}}\chi(\alpha_p) \, , \qquad
\alpha_p = \frac{2\pi p}{n} \, .
\label{dSnint}
\ee
Since the relation is linear in the generating function, one can directly apply it to the difference,
and using \eqref{dchi} one arrives at
\be
\delta S_n = \frac{1}{1-n}\sum^{\frac{n-1}{2}}_{p = -\frac{n-1}{2}}\ln \cos \left(\frac{\pi p r}{n}\right).
\label{dSnsum}
\ee

\begin{figure}[htb]
\center
\includegraphics[width=0.6\textwidth]{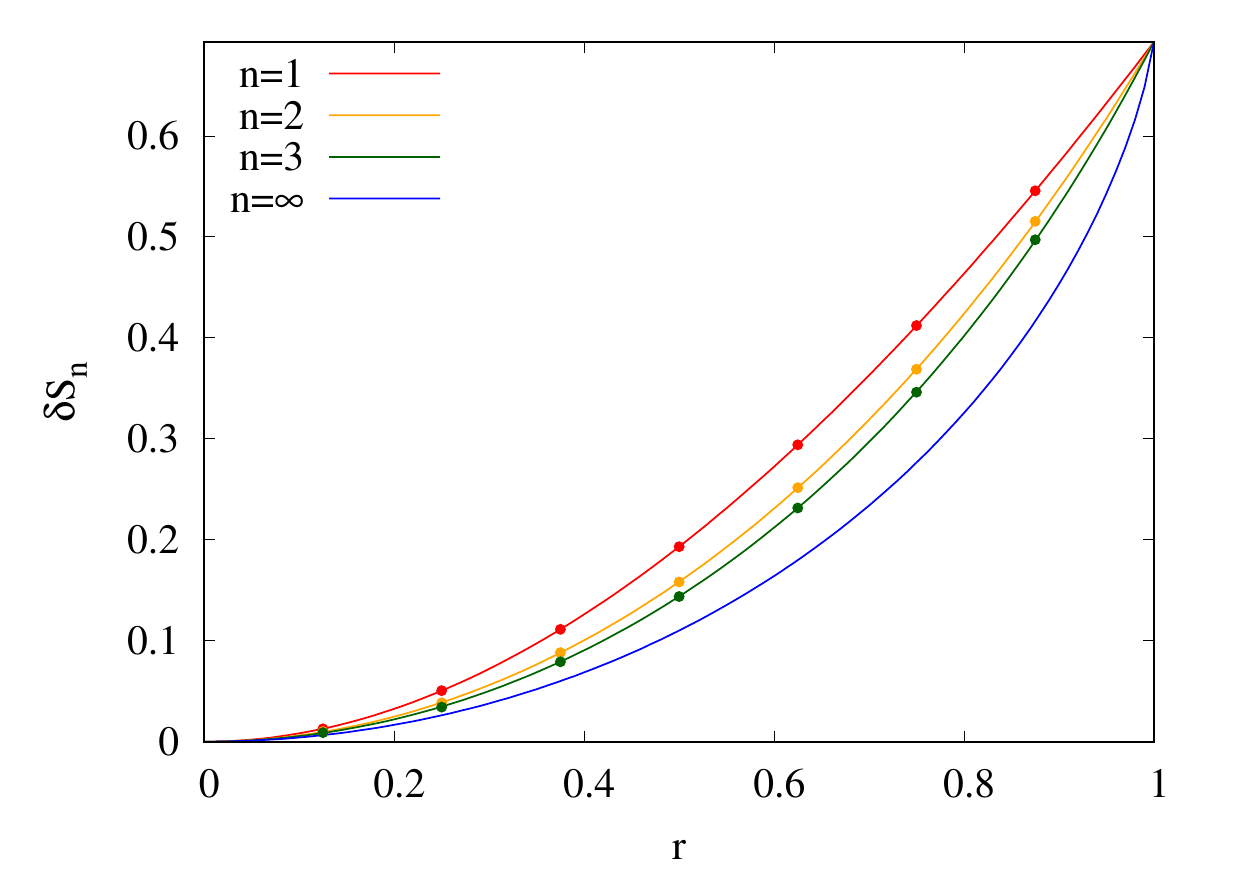}
\caption{Zero-mode R\'enyi entropies $\delta S_n$ as a function of the ratio $r$ for various $n$.
The symbols of matching color show the numerical results obtained via data fits to \eqref{dSfit}.}
\label{fig:dSnhom}
\end{figure}

The zero-mode R\'enyi entropies are shown in Fig.~\ref{fig:dSnhom} and compared against
numerical calculations, performed with a fixed ratio $r=\ell/(2L)$ and increasing $L$. One should
note, that the numerical data shows relatively strong finite-size corrections, and the entropy
difference is well described by
\eq{
\delta S_n(\ell,L) = \delta S_n(r) + (-1)^\ell\frac{a}{\ell^{1/n}} \, .
\label{dSfit}}
One has thus an alternation with the parity of the subsystem size and some unusual correction scaling with a power $1/n$, that originates from the pure state and was noticed earlier \cite{FC11}.
We used the above ansatz to fit the data and extract the scaling part $\delta S_n(r)$. As observed in Fig.~\ref{fig:dSnhom},
the fits are in excellent agreement with the analytical results. In general, the curves for each $n$
interpolate smoothly and monotonically between the values $0$ and $\ln(2)$. A special value for $r=1/2$ is $\delta S= \ln(2)-1/2$. Note that the line for $n=1$ is obtained by a numerical evaluation of the integral \eqref{dShom}, whereas the
$n=\infty$ case follows from converting the sum \eqref{dSnsum} into an integral.

%%%%%%%%%%%%%%%%%%%%%%%%%%%%%%%
%%%%%%%%%%%%%%%%%%%%%%%%%%%%%%%

\subsection{Spectral shift}

The agreement between the analytical and numerical results is remarkable, despite the fact
that the actual numerical spectra are still very far from being continuous and densely spaced
for the chain sizes considered. To better understand the mechanism behind the emergence
of the zero-mode contribution, we shall have a closer look at the entanglement spectra.
According to \eqref{rhoeps}, the density of states is obtained as the derivative of the
spectral function $\kappa(\varepsilon)$, which is simply the inverted spectrum
$\varepsilon_\kappa$ plotted against the integer index $\kappa$. Thus the spectral
function $\kappa(\varepsilon)$ simply counts the number of eigenvalues up to $\varepsilon$.

For the case of a pure Fermi sea, the asymptotics of such spectral functions is known
for the infinite or semi-infinite hopping chain \cite{EP13}. These results are based on the
analysis of the discrete sine kernel from the original work of Slepian \cite{Slepian78}.
Although the correlation matrix $C_1$ in \eqref{C1hom} is kind of a deformed sine kernel,
we are not aware of any rigorous results for its spectral function $\kappa_1(\varepsilon)$.
Nevertheless, we try to guess the result by analogy to \cite{Slepian78}, as well as from the relation
of $\kappa_1(\varepsilon)$ to the entropy. Namely, for the half-filled ground state $\hat \rho_1$
we put forward the ansatz
\eq{
\kappa_1-\bar\kappa_1=\frac{\varepsilon}{2\pi^2} \ln\Big(\frac{8L}{\pi}\sin(\pi r)\Big)-
\frac{1}{\pi}\varphi \Big(\frac{\varepsilon}{2\pi}\Big),
\label{k1eps}}
where $\bar\kappa_1=\Tr(\hat\rho_1 \hat N_A)+1/2$ is just a constant related to the average number of particles in $A$,
while $\varphi(z)$ is given via the Gamma function as
\eq{
\varphi(z)=\mathrm{arg\,}\Gamma(1/2+iz) \, .
}
Calculating now the entropy with the corresponding density of states one has
\eq{
S_1 = \int_{-\infty}^{\infty} \dd \varepsilon \,
\frac{\dd \kappa_1}{\dd\varepsilon} s(\varepsilon)=
\frac{1}{6}\ln\left(\frac{8L}{\pi}\sin(\pi r)\right) + \frac{\mathcal{C}}{2},
\label{S1hom}}
where the constant term follows from the non-linear part of the spectral function as
\eq{
\mathcal{C} = -\frac{1}{\pi^2}\int_{-\infty}^{\infty} \dd \varepsilon \, s(\varepsilon)
\varphi'\left(\frac{\varepsilon}{2\pi}\right)\approx 0.495 \, .
\label{C}}

\begin{figure}[htb]
\center
\includegraphics[width=0.6\textwidth]{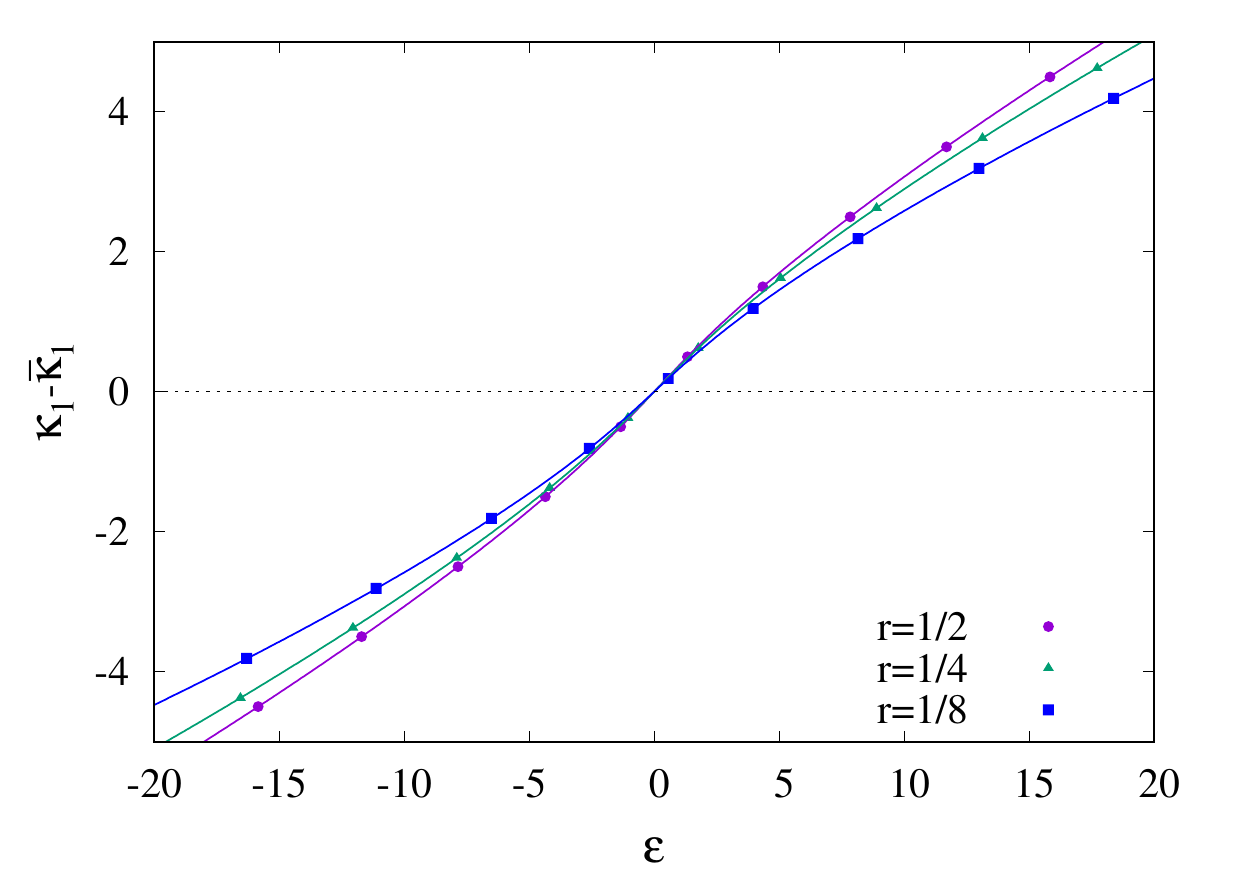}
\caption{Spectral function $\kappa_1(\varepsilon)$ of the Fermi sea ground state for various
ratios $r$ and $L=100$. The solid lines of matching color correspond to the ansatz \eqref{k1eps}.}
\label{fig:k1eps}
\end{figure}

The expression \eqref{S1hom} resembles very closely the result for the open chain
without boundary fields, which was studied in \cite{FC11}. Indeed, the only modification
is that $2(4L+2)$ is replaced by $8L$ in the argument of the logarithm. This is motivated
by the fact, that the eigenfunctions of the simple open chain vanish at sites $m=0$ and
$m=2L+1$, and one thus needs to add two extra sites to embed the chain and its
mirror image into a periodic ring \cite{FC11}. Here, instead, the eigenfunctions \eqref{phihom}
vanish at $m=1/2$ and $2L+1/2$, and one can argue that the two extra sites are not needed
and the effective length of the corresponding ring is $4L$. Note also, that the constant
$\mathcal{C}$ in \eqref{C} is precisely the one that enters the entropy of the ring \cite{JK04},
which now appears with a factor $1/2$ due to the single boundary between the subsystem
$A$ and the rest of the chain. After having motivated our ansatz \eqref{k1eps} for the spectral
function, we now compare it against exact numerical calculations in Fig.~\ref{fig:k1eps} for
various ratios $r$, and observe an excellent agreement.

%%%%%%%%%%%%%%%%%%%%%%%%%%%%%%%%%%%%%%%%%
%%%%%%%%%%%%%%%%%%%%%%%%%%%%%%%%%%%%%%%%%

\begin{figure}[htb]
\center
\includegraphics[width=0.6\textwidth]{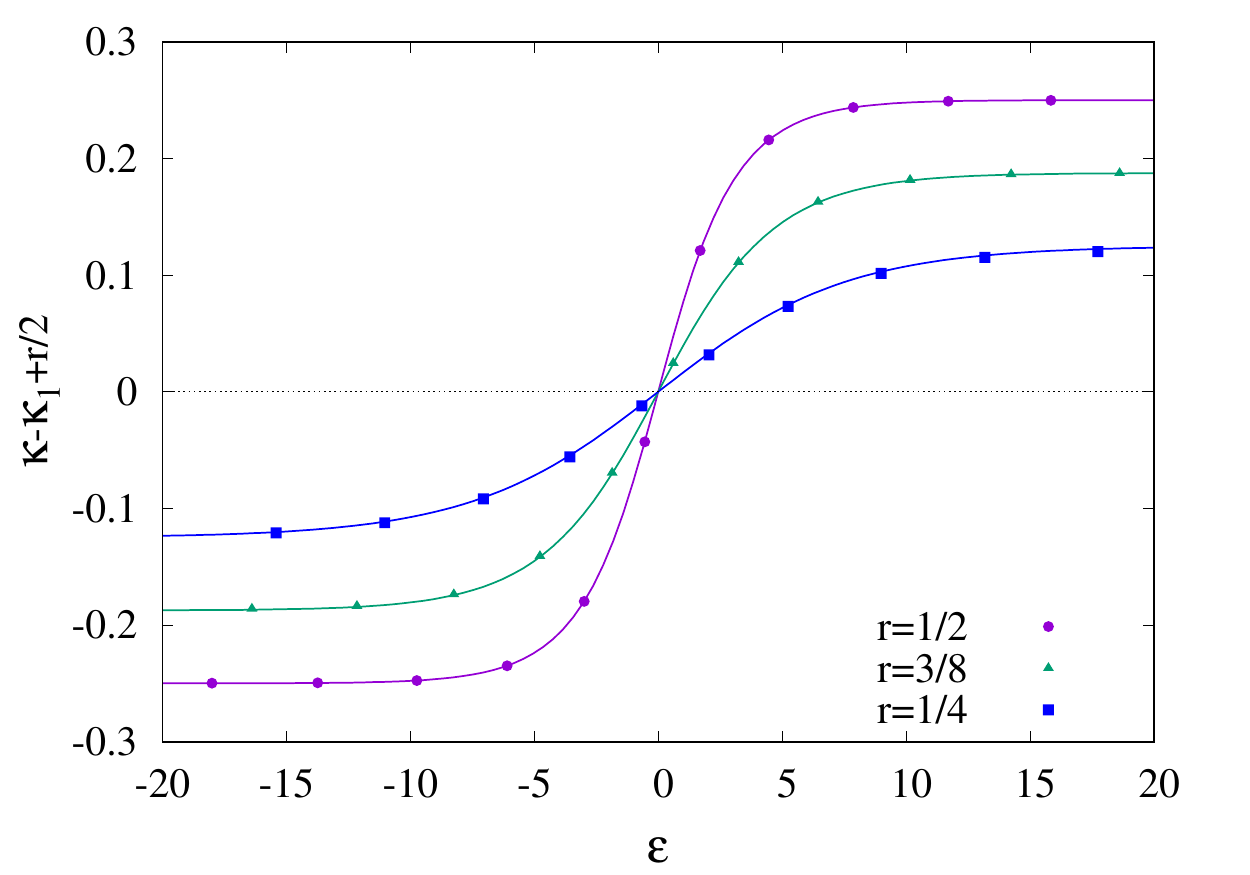}
\caption{Spectral shift for various $r$ and $L=100$. The solid lines show the
analytical result \eqref{dkepshom}.}
\label{fig:dkepshom}
\end{figure}

We now move forward to study the spectral function $\kappa(\varepsilon)$ of the mixed state,
associated to the correlation matrix $C_A$ in \eqref{Chom}. This is related to the matrix $C_{1,A}$
by a rank-one update, which induces slight shifts between the eigenvalues 
$\varepsilon_\kappa$ and $\varepsilon_{1,\kappa}$ in the corresponding spectra.
It should be stressed, however, that in the numerics the independent variable $\kappa$ is always an integer.
To extract the spectral shift $\delta\kappa(\varepsilon)=\kappa(\varepsilon)-\kappa_1(\varepsilon)$,
we have to treat $\varepsilon$ as the independent variable, i.e. we invert $\varepsilon_\kappa \to \kappa(\varepsilon_\kappa)$
and subtract $\kappa_1(\varepsilon_\kappa)$ by using our ansatz \eqref{k1eps}.
In the thermodynamic limit, the spectral shift follows by integrating
the density $\delta\rho(\varepsilon)$ in \eqref{drhoeps}, which yields
\eq{
\delta\kappa(\varepsilon) =
\frac{1}{\pi} \arctan
\left[\tan \left(\frac{\pi r}{2}\right)\tanh \left(\frac{\varepsilon r}{2}\right)\right],
\label{dkepshom}}
where the integration constant was chosen such that $\delta\kappa(0)=0$.
The comparison of the numerically extracted spectral shift to the analytical result
\eqref{dkepshom} is shown in Fig. \ref{fig:dkepshom}, with a remarkable agreement.
Note that, in order to bring the numerical data to a symmetric form, a constant $r/2$
has been added, which exactly corresponds to the difference $\bar\kappa_1-\bar\kappa$
of the average particle number in $A$.

%%%%%%%%%%%%%%%%%%%%%%%%%%%%%%%%%%%%%%%%%
%%%%%%%%%%%%%%%%%%%%%%%%%%%%%%%%%%%%%%%%%

To conclude this section we mention, that the zero-mode contributions can also be
investigated for a periodic ring of size $2L$, with boundary conditions
$H_{1,2L}=H_{2L,1}=-1/2$ in \eqref{H}. The momenta are then quantized as $\pi k/L$ with
$k=-L+1, \dots,L$, and one has a pair of zero modes with $\pm k_F=L/2$ for even half-chain sizes.
Both of them have to be included with an occupation $n_{\pm k_F}=1/2$, which
leads to a mixed ground state composed of four different pure states.
The ansatz \eqref{k1eps} has to be modified by replacing $8L \to 4L$ and multiplying the r.h.s.
with a factor two, that accounts for the increased density of states due to the two boundary points of $A$.
Carrying out the numerical analysis analogously, one finds that the spectral shift $\delta\kappa(\varepsilon)$
as well as the resulting entropy difference $\delta S$ are both multiplied by a factor of two.
Since the results look very similar to the ones obtained for the open chain in this section, we do not report them here.

\section{Zero mode for the defect}\label{sec:Defect}

In the previous section we have presented an alternative derivation of the zero-mode contribution, that has been studied previously for a translational invariant system \cite{KVW17}. We shall now extend these results for free-fermion chains with a particular form of a defect, with the nonzero elements of the hopping matrix given by
\eq{
H'_{m,m+1}= H'_{m+1,m} = \left\{
\begin{array}{ll}
-1 & \quad m \ne L \\
-\lambda  & \quad m = L
\end{array}\right. , \qquad
H'_{L,L} = -H'_{L+1,L+1}=\sqrt{1-\lambda^2} \, ,
\label{Hdef}}
while the boundary potentials $H'_{1,1}=H'_{2L,2L}=\frac{1}{2}$ are the same
as in the homogeneous case. Note that we use a prime notation to distinguish
the quantities defined for the defect problem from the ones of the homogeneous
case with $\lambda=1$. In fact, it turns out that the eigenmodes of the two
Hamiltonians are intimately related via
\eq{
\omega'_k =\omega_k = -\cos \Big(\frac{\pi k}{2L}\Big)  \, , \qquad
\phi'_k(m) = \left\{
\begin{array}{ll}
\alpha_k \phi_k(m) & \quad m \le L \\
\beta_k \phi_k(m) & \quad  m > L
\end{array}\right. .
\label{phikdef}
}
In other words, the spectra are identical, supporting a zero mode at $k=k_F=L$ for \emph{arbitrary}
values of $\lambda$, while the eigenvectors are related to their homogeneous counterparts
\eqref{phihom} by a rescaling that is different on the left/right hand side of the defect and is
given by the factors
\eq{
\alpha_k^2 = 1 + (-1)^k \sqrt{1-\lambda^2}, \qquad
\beta_k^2 = 1 - (-1)^k \sqrt{1-\lambda^2} \, .
\label{ab}
}

The situation is thus completely analogous to the case of a simple open chain, where
this so-called conformal defect was studied previously \cite{EP12}. The terminology derives
from the fact that the transmission amplitude $s=\lambda$ is independent of the incoming
momentum $k$. Furthermore, for the particular case of a half-chain bipartition,
the entanglement spectra are related as
\eq{
\cosh \Big(\frac{\varepsilon'_{\sigma,\kappa}}{2}\Big)=
\frac{1}{\lambda}\cosh \Big(\frac{\varepsilon_{\sigma,\kappa}}{2}\Big) \, ,
\label{epsdef}}
where $\sigma=0,1$ refers to the pure states in \eqref{01}. The relation is proved in appendix \ref{app:epsdef}.
%by considering the overlap matrices $A'_{k,l}$ and $A_{k,l}$ of the filled modes, which are known to be isospectral to $C'_{m,n}$ and $C_{m,n}$, up to trivial eigenvalues $0$ or $1$.
One should stress that \eqref{epsdef} holds for arbitrary particle
number $N$, but only for a half-chain $A=\left[1,L\right]$, we thus restrict our attention to this case.
Importantly, the main feature of the entanglement spectra for the defect is the presence of a gap between
\eq{
\varepsilon_\pm = \pm 2 \, \mathrm{acosh}(\lambda^{-1})
\label{epspm}}
where no eigenvalues are allowed.

Before turning to the FCS, let us comment on a special property of the spectra that will become
important for the defect. Indeed, it turns out that the homogeneous spectrum at $r=1/2$ has a particle-hole symmetry, which leads to parity effects in terms of the particle number $N$. For an even $L$, this implies that the spectra
of the state with $N=L$ contain $N/2$ pairs with $\pm \varepsilon_{1,\kappa}$ and
$\pm \varepsilon'_{1,\kappa}$, respectively. 
On the other hand, for $N=L-1$ particle-hole symmetry implies the existence of
a single $\varepsilon_{0,\kappa}=0$, with the rest of the nonzero eigenvalues coming in pairs.
One then observes that the corresponding defect eigenvalue is located on the upper edge
$\varepsilon_+$ of the gap, thus making the spectrum slightly asymmetric.
Note that an eigenvalue on the lower edge $\varepsilon_-$ of the gap appears when considering
the spectra for the right half-chain. For odd $L$ the two cases discussed above are simply interchanged.

We are now ready to derive the formula analogous to \eqref{detratio}.
Rewriting the FCS generators $\chi'_\sigma(\alpha)$ with $\sigma=0,1$ in the pure states \eqref{01}
in terms of the corresponding spectra $\varepsilon'_{\sigma,\kappa}$, one has
\eq{
\mathrm{Re} \, \chi'_\sigma(\alpha) = \frac{1}{2}\sum_{\kappa} \ln
\left[1-\frac{\sin^2(\frac{\alpha}{2})}{\cosh^2(\frac{\varepsilon'_{\sigma,\kappa}}{2})}\right],
\qquad
\, \mathrm{Im} \, \chi'_\sigma(\alpha) = \frac{1}{2i} \sum_{\kappa} \ln
\left[\frac{\ex^{\varepsilon'_\kappa} +\ex^{i\alpha}}{\ex^{\varepsilon'_\kappa} +\ex^{-i\alpha}}\right].
}
The real part can immediately be related via \eqref{epsdef} to the homogeneous FCS
\eq{
\mathrm{Re} \, \chi_\sigma'(\alpha) = \mathrm{Re} \, \chi_\sigma(\alpha') \, , \qquad
\sin \Big(\frac{\alpha'}{2}\Big) = \lambda \sin \Big(\frac{\alpha}{2}\Big)
}
 by a change of variables $\alpha\to\alpha'$. Taking the logarithm of \eqref{detratio}
implies then $\mathrm{Re} \left[ \chi'_1(\alpha)- \chi'_0(\alpha) \right]$=0.
Dealing with the imaginary part requires some care due to the parity effects discussed above.
For a perfectly symmetric spectrum, one can simply add the contributions from the pairs.
Using
\eq{
\frac{\ex^{\varepsilon'_{1,\kappa}} +\ex^{i\alpha}}{\ex^{\varepsilon'_{1,\kappa}} +\ex^{-i\alpha}}
\frac{\ex^{-\varepsilon'_{1,\kappa}} +\ex^{i\alpha}}{\ex^{-\varepsilon'_{1,\kappa}} +\ex^{-i\alpha}}=
\ex^{2i\alpha} \, ,
}
this gives for an even $L$
\eq{
\mathrm{Im} \, \chi'_1(\alpha) = \alpha \frac{L}{2} \, .
%\frac{1}{2} \sum_{\kappa} \ln
%\left[\frac{\ex^{\varepsilon'_\kappa} +\ex^{i\alpha}}{\ex^{\varepsilon'_\kappa} +\ex^{-i\alpha}}\right]
}
For $N=L-1$, the extra eigenvalue sitting at the gap edge has to be added separately
\eq{
\mathrm{Im} \, \chi'_0(\alpha) =
\alpha \Big(\frac{L}{2}-1\Big)+\frac{1}{2i} \mathrm{Im} \ln
\left[\frac{\ex^{\varepsilon_+} +\ex^{i\alpha}}{\ex^{\varepsilon_+} +\ex^{-i\alpha}}\right].
}
Inserting \eqref{epspm} and using some identities, the difference can be expressed as
\eq{
\mathrm{Im} \left[ \chi'_1(\alpha)- \chi'_0(\alpha) \right]=
\frac{\alpha}{2} +
\tan^{-1}\left(\sqrt{1-\lambda^2}\tan\Big(\frac{\alpha}{2}\Big)\right),
\label{imdchi}}
which is the analogue of \eqref{detratio} in the presence of the defect.
The case of odd $L$ is straightforward to obtain, and gives a result similar to \eqref{imdchi}
but with a minus sign in front of the second term.

It should be noted, that the above result implies a highly nontrivial change in the FCS. Contrary to the homogeneous case, it is not a simple shift in the average particle number, but affects higher order cumulants as well. In particular, since \eqref{imdchi} is an odd function of $\alpha$, the FCS becomes skewed.
In turn, it yields for the difference of the FCS between the mixed and pure states
\eq{
\delta \chi_\pm(\alpha) = \chi'(\alpha)-\chi'_{0}(\alpha) =
\ln \left[\frac{1+\ex^{i \alpha/2 \pm i
\tan^{-1}(\sqrt{1-\lambda^2}\tan(\alpha/2))}}{2}\right],
\label{dchidef}}
where the $\pm$ sign refers to $L$ being even or odd, respectively. It is instructive to check the limit $\lambda=1$, where $\delta \chi_\pm(\alpha)=\delta \chi(\alpha)$ reproduces the homogeneous result \eqref{dchi} with $r=1/2$. On the other hand, for a disconnected chain $\lambda=0$ one has the simple expressions
\be
\delta \chi_+(\alpha) =  \ln \left( \frac{1+e^{i\alpha}}{2}\right), \qquad \delta \chi_-(\alpha) =  0,
\ee
which correspond again to Eq.~\eqref{dchi} with the effective ratios $r=1,0$, respectively. This immediately
yields the values $\delta S_e = \ln(2)$ and $\delta S_o=0$ for the even/odd case in the limit $\lambda \to 0$.

\subsection{Spectral density}

With the result \eqref{dchidef} at hand, we can now perform the calculation for the spectral density in the exact same way as for the homogeneous case. Namely, we
introduce the variable $z$ in \eqref{z}, and employ the identities
\be
e^{i\alpha/2} = \sqrt{1-\frac{1}{z}}, \qquad \tan(\alpha/2) = \frac{1}{i}\frac{1}{1-2z}, \qquad e^{\tanh^{-1}(x)} = \sqrt{\frac{1+x}{1-x}}.
\ee
%
%We now consider $\lambda \in (0,1)$
Furthermore, we also introduce the parameters
\be
z_{\pm} = \frac{1\pm \sqrt{1-\lambda^2}}{2},
\ee
such that \eqref{dchidef} can be rewritten as
\be
\delta \chi_\pm(z) = \ln \left( \frac{1+\sqrt{1-\frac{1}{z}} \sqrt{\frac{z-z_\pm}{z-z_\mp}}}{2} \right).
\ee
Rearranging the expression one gets
\be\label{eq:dChi_defect}
\delta \chi_\pm(z) = \ln \left(\frac{ \sqrt{z(z_\mp-z)} + \sqrt{(1-z)(z-z_\pm)} }{2} \right) - \frac{1}{2}\ln\left( z(z-z_\mp)\right),
\ee
where the first term is responsible for the continuous part of the spectral density, while the second one simply contributes the delta functions $-\frac{1}{2}(\delta(z)+\delta(z-z_\mp))$. Using
\be
\frac{d}{dz} \sqrt{(z-a)(b-z)} = \frac{\frac{a+b}{2}-z}{\sqrt{(z-a)(b-z)}},
\ee
we get for the derivative of the first term in \eqref{eq:dChi_defect} after tedious but straightforward algebra
\be\label{eq:der_Chi}
%\begin{split}
%&\frac{d}{dz}\ln \left(\frac{ \sqrt{z(z_\mp-z)} + \sqrt{(1-z)(z-z_\pm)} }{2} \right) = \\
\frac{1}{z_\pm(1-2z)}\left( -z_\pm 
 -\left(  \frac{z_\mp}{2}-z\right)\frac{\sqrt{(1-z)(z-z_\pm)}}{\sqrt{z(z_\mp -z)}} + \left(  \frac{z_\pm +1}{2}-z\right)\frac{\sqrt{z(z_\mp -z)}}{\sqrt{(1-z)(z-z_\pm)}} \right).
%\end{split}
\ee

At this point, one should be particularly careful to deal properly with the branch cuts appearing in the region $z \in (0,z_-)\cup(z_+,1)$, due to the presence of the square roots. We first consider the case $z\in (0,z_-)$, where the jump across the branch cut is
\be
\lim_{\epsilon \to 0^+} \sqrt{(z\mp i\epsilon -z_\pm)(1-z\pm i\epsilon)} = \mp i\sqrt{(1-z)(z_\pm-z)},
\ee
and thus we arrive at
\be\label{eq:deltaRo_c}
\delta \rho_{\pm}(\zeta) =  \frac{1}{\pi z_\pm (1-2\zeta)}\left( \left(\frac{z_\mp}{2}-\zeta\right) \sqrt{\frac{(1-\zeta)(z_\pm-\zeta)}{\zeta(z_\mp-\zeta)}} + \left( \frac{z_\pm +1}{2}-\zeta\right)\sqrt{\frac{\zeta(z_\mp -\zeta)}{(1-\zeta)(z_\pm-\zeta)}} \right).
\ee
%\be\label{eq:deltaRo_c}
%\delta \rho_{\pm}(z) =  \frac{1}{\pi z_\pm (1-2z)}\left( \left(\frac{z_\mp}{2}-z\right) \sqrt{\frac{(1-z)(z_\pm-z)}{z(z_\mp-z)}} + \left( \frac{z_\pm +1}{2}-z\right)\sqrt{\frac{z(z_\mp -z)}{(1-z)(z_\pm-z)}} \right).
%\ee
For $z\in(z_+,1)$ one gets the same result, and thus Eq. \eqref{eq:deltaRo_c} provides the continuous part of the spectral density, which turns out to be symmetric under $\zeta \rightarrow 1-\zeta$. Further singular contributions to the spectral density are delivered by the poles of the resolvent. In particular, the term $(1-2z)$ in the denominator of Eq. \eqref{eq:der_Chi} might give rise to a pole at $z=1/2$. The presence of the pole can be spotted by approximating \eqref{eq:der_Chi} around $z=1/2$, where we find the limits
\be
\delta R_+(z)(z-1/2) \simeq 1, \qquad \delta R_-(z)(z-1/2) \simeq 0.
\ee
The pole is thus present only for $\delta R_+(z)$, leading to the even/odd spectral densities
\be
\delta \rho_e(\zeta) = \delta \rho_+(\zeta) -\frac{1}{2}\delta(\zeta-z_-)+\delta(\zeta-1/2), \qquad \delta \rho_o(\zeta) = \delta \rho_-(\zeta) -\frac{1}{2}\delta(\zeta-z_+),
\ee
where we discarded again the irrelevant delta functions at $\zeta=0$.

\begin{figure}[htb]
\center
\includegraphics[width=0.6\textwidth]{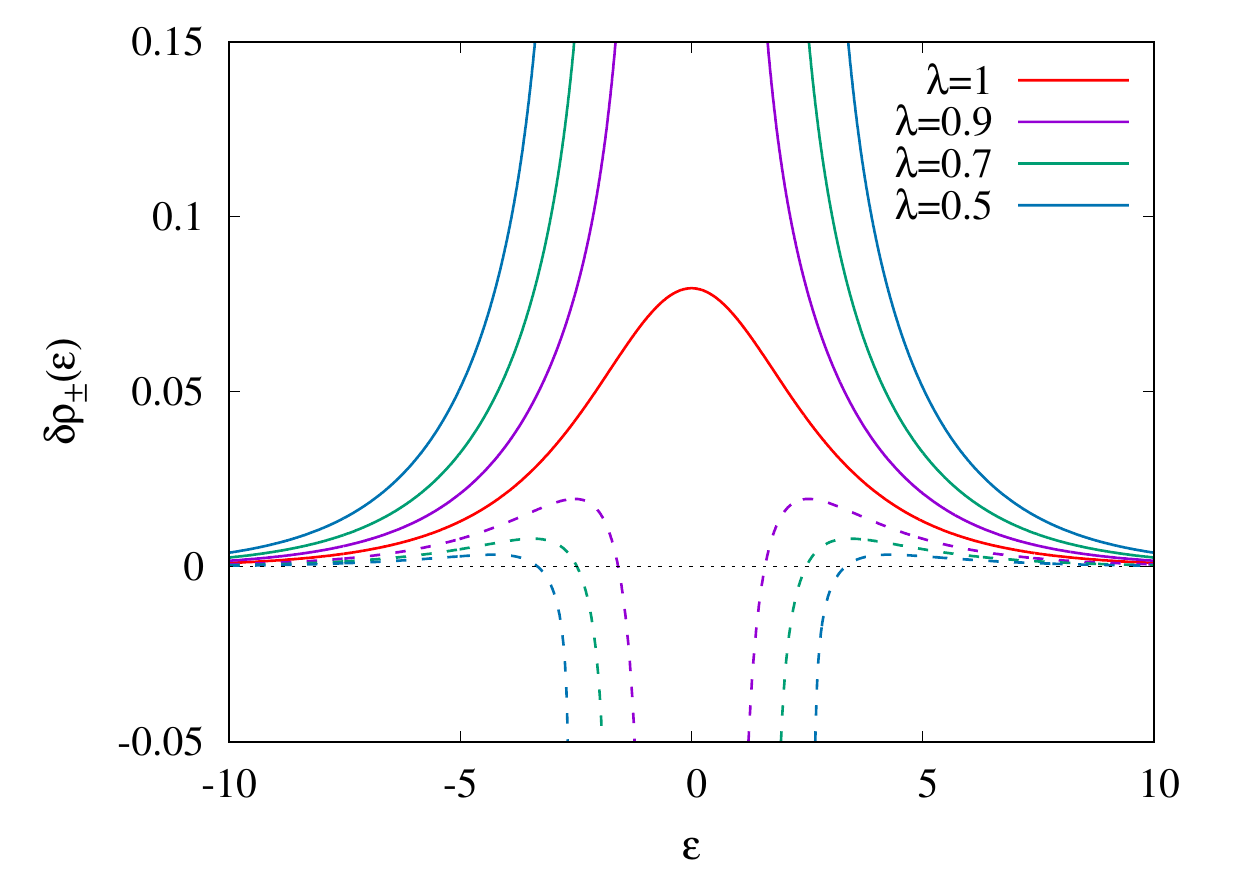}
\caption{Continuous part of the spectral density $\delta\rho_+(\varepsilon)$
(dashed lines) and $\delta\rho_-(\varepsilon)$ (solid lines) for various values of $\lambda$.}
\label{fig:drhodef}
\end{figure}

The change in the density of the entanglement spectrum
is obtained by transforming variables via \eqref{rhoepszeta}. Noting that $z_\pm=\zeta(\varepsilon_\mp)$, after some algebra one finds
\eq{
\delta\rho_e(\varepsilon)=
\delta\rho_+(\varepsilon)+\delta(\varepsilon)-\frac{1}{2}\delta(\varepsilon-\varepsilon_+) \, ,
\qquad
\delta\rho_o(\varepsilon)=\delta\rho_-(\varepsilon) -\frac{1}{2}\delta(\varepsilon-\varepsilon_-) \, ,
\label{drhoeo}}
where the continuous part of the density is given by
\eq{
\delta\rho_\pm(\varepsilon) = 
\frac{1}{2\pi(1 \pm \sqrt{1-\lambda^2}) \, |\sinh (\varepsilon)|}\left(
\sqrt{\lambda^2 \cosh^2 \left(\frac{\varepsilon}{2}\right)-1} \mp
\frac{\sqrt{1-\lambda^2}}{\sqrt{\lambda^2 \cosh^2 \left(\frac{\varepsilon}{2}\right)-1}}
%\frac{\lambda^2 \cosh^2 \left(\frac{\varepsilon'}{2}\right)-1\pm \sqrt{1-\lambda^2}}
%{\sqrt{\lambda^2 \cosh^2 \left(\frac{\varepsilon'}{2}\right)-1}}
\right).
\label{drhopm}}
These are shown in Fig.~\ref{fig:drhodef} for some values of $\lambda$, with the dashed
and solid lines corresponding to $\delta\rho_+(\varepsilon)$ and $\delta\rho_-(\varepsilon)$, respectively.
As is clear from \eqref{drhopm}, both functions diverge around $\varepsilon_\pm$, and they move away
from the homogeneous ($\lambda=1$) curve in opposite directions. In fact, the density difference
$\delta\rho_+(\varepsilon)$ is negative close to the gap edges, and it becomes very
small moving away from it. Hence the dominant contribution to the even spectral density is actually
delivered by the delta peaks in \eqref{drhoeo} for smaller values of $\lambda$.

%%%%%%%%%%%%%%%%%%%%%%%%%%%%%
%%%%%%%%%%%%%%%%%%%%%%%%%%%%%

\subsection{Numerical results}

\begin{figure}[htb]
\center
\includegraphics[width=0.6\textwidth]{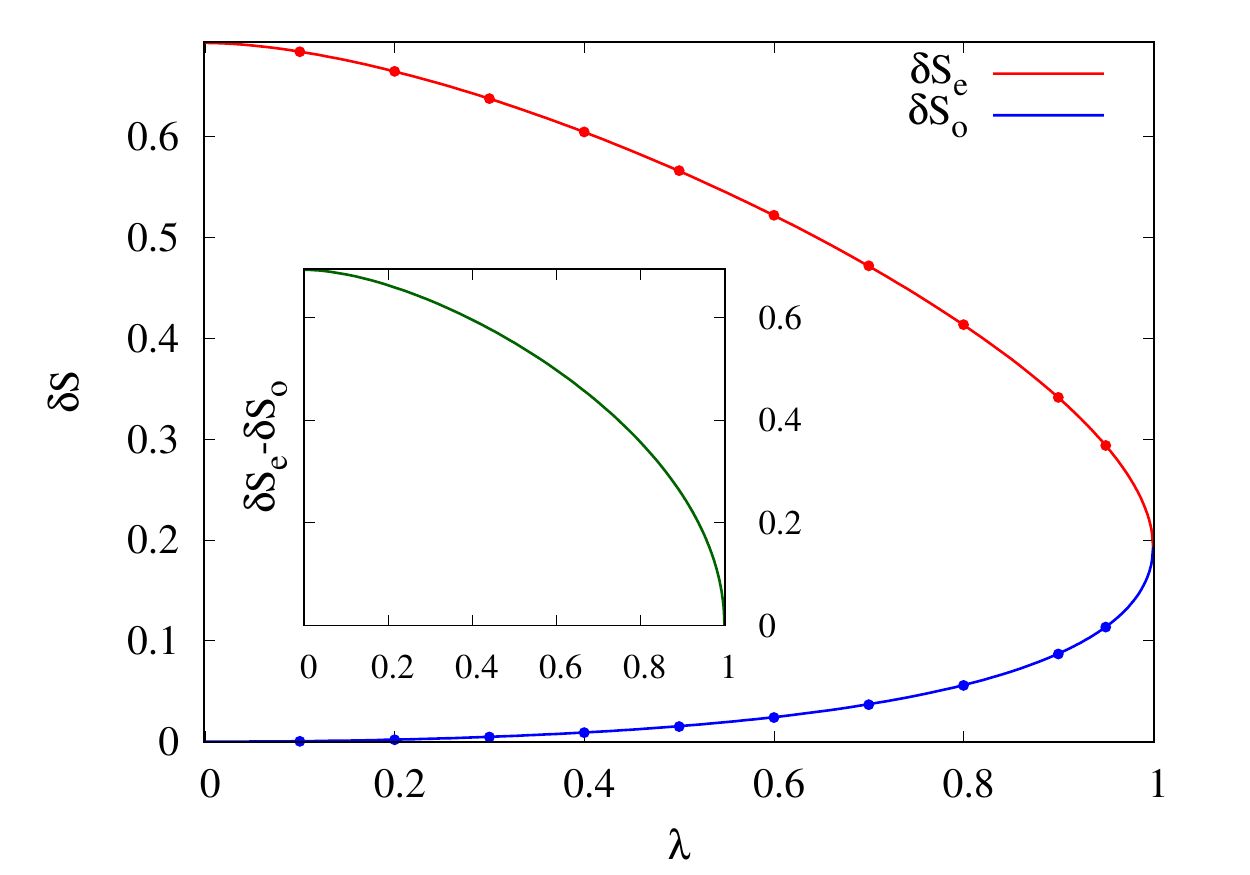}
\caption{Zero-mode entropy $\delta S_e$ and $\delta S_o$ for even/odd half-chain lengths
as a function of $\lambda$. The symbols show the fits to the numerical data,
and the lines correspond to the analytical result in \eqref{dSeo}. The inset shows the
difference $\delta S_e-\delta S_o$.}
\label{fig:dSdef}
\end{figure}

In the following we present our numerical results for the defect, and compare them to the analytical predictions obtained via the spectral densities in \eqref{drhoeo}. We start with the entanglement entropy, focusing on the case $n=1$. The zero-mode entropy for the defect follows as
\eq{
\delta S_e = 2\int_{\varepsilon_+}^{\infty} \dd \varepsilon \, 
\delta\rho_+(\varepsilon) \, s(\varepsilon) -\frac{s(\varepsilon_+)}{2}+\ln(2) \, ,
\qquad
\delta S_o = 2\int_{\varepsilon_+}^{\infty} \dd \varepsilon \, 
\delta\rho_-(\varepsilon) \, s(\varepsilon) -\frac{s(\varepsilon_+)}{2} \, ,
\qquad
\label{dSeo}}
where we used the symmetry of the spectral and entropy densities under $\varepsilon\to-\varepsilon$.
In Fig.~\eqref{fig:dSdef} we compare the above integral expressions to the entropy difference $S-S_1$
obtained from the numerics. As expected, the data converge to different values for even/odd $L$, and shows finite size corrections of the form
$\delta S_{e,o}(L)=\delta S_{e,o}+a_{e,o}/L$.
Fitting this expression for increasing $L$,
one obtains the values indicated by the symbols in Fig.~\ref{fig:dSdef}. They show an excellent agreement with the analytical results. Note that the zero-mode
entropy $\delta S_e$ ($\delta S_o$) increases (decreases)
monotonously as the defect strength goes towards $\lambda\to 0$. Their difference $\delta S_e-\delta S_o$
is shown in the inset.

%%%%%%%%%%%%%%%%%%%%%%%%%%%%%
%%%%%%%%%%%%%%%%%%%%%%%%%%%%%

\begin{figure}[htb]
\center
\includegraphics[width=0.49\textwidth]{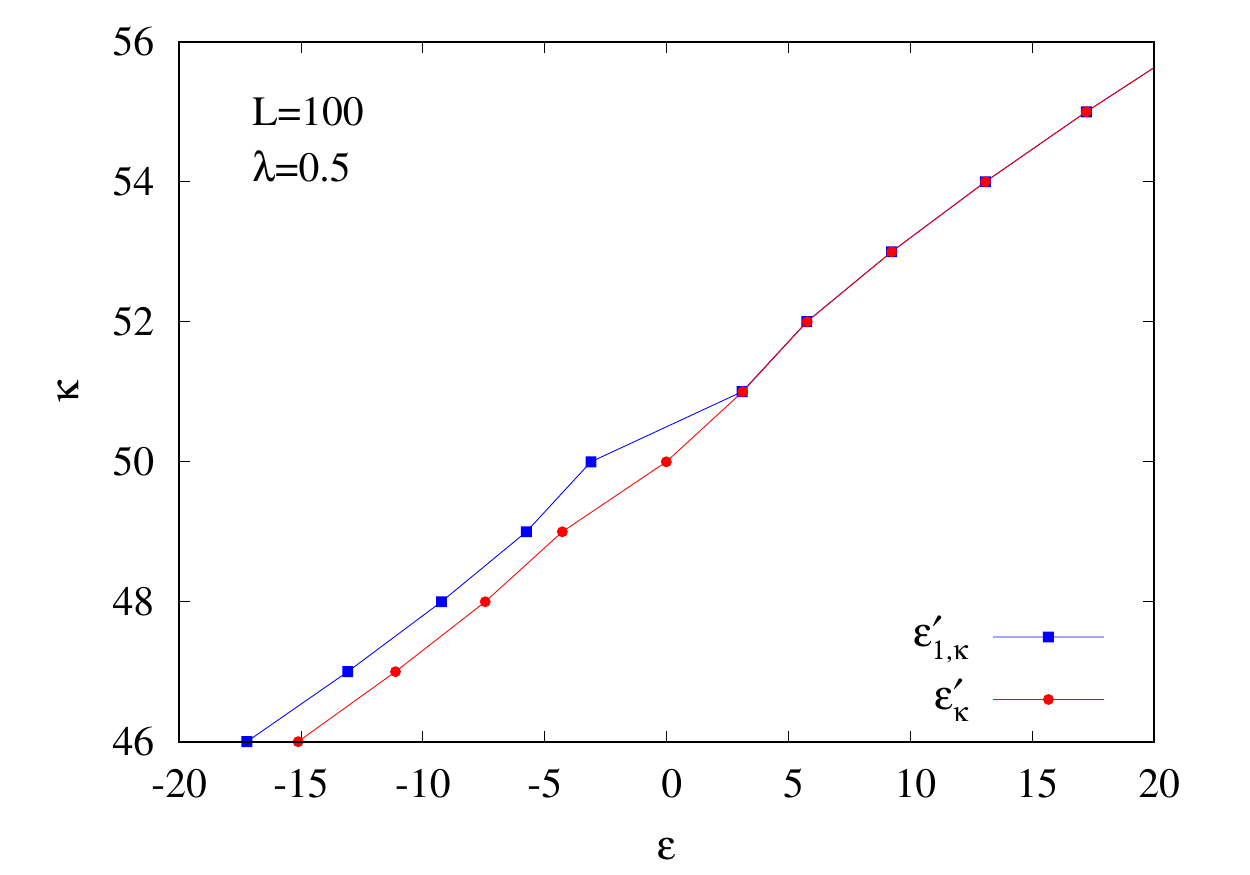}
\includegraphics[width=0.49\textwidth]{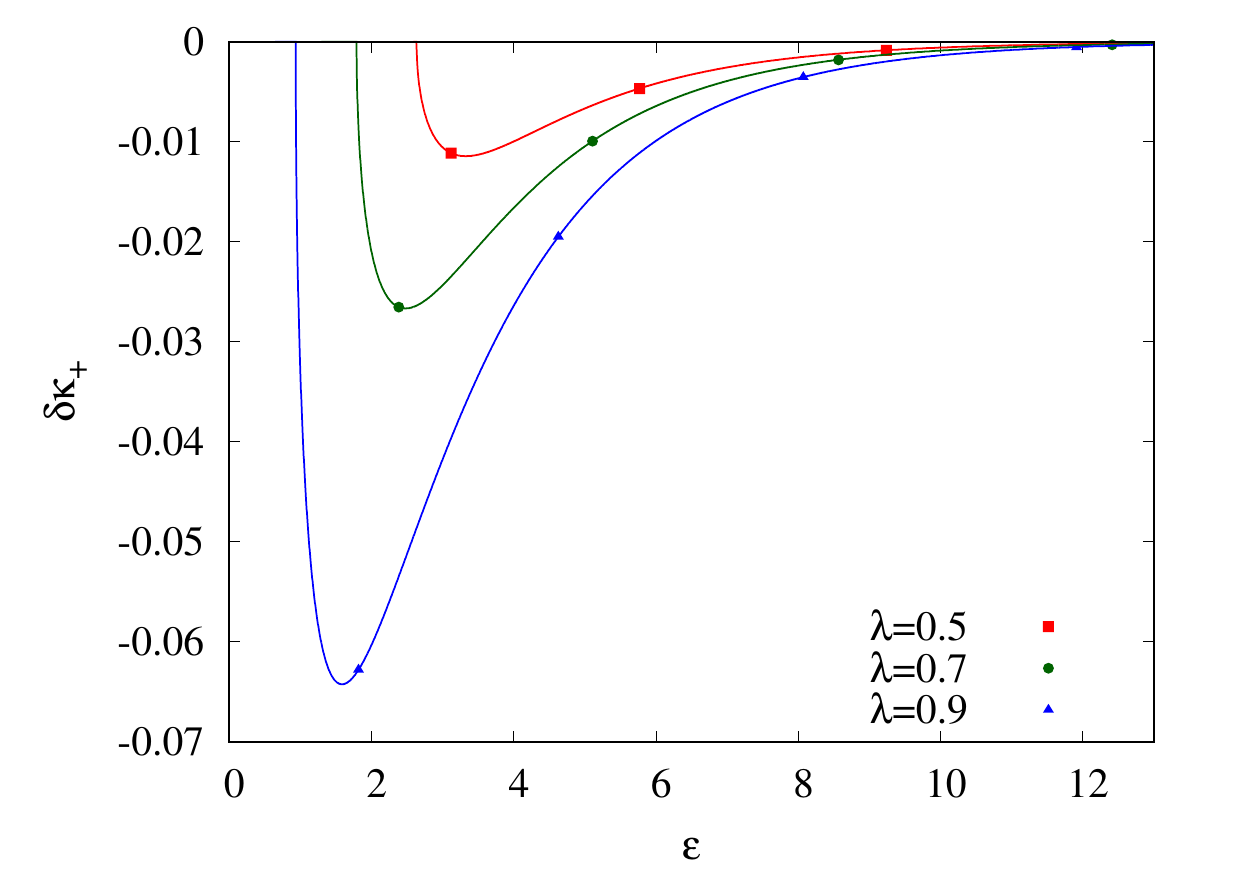}
\includegraphics[width=0.49\textwidth]{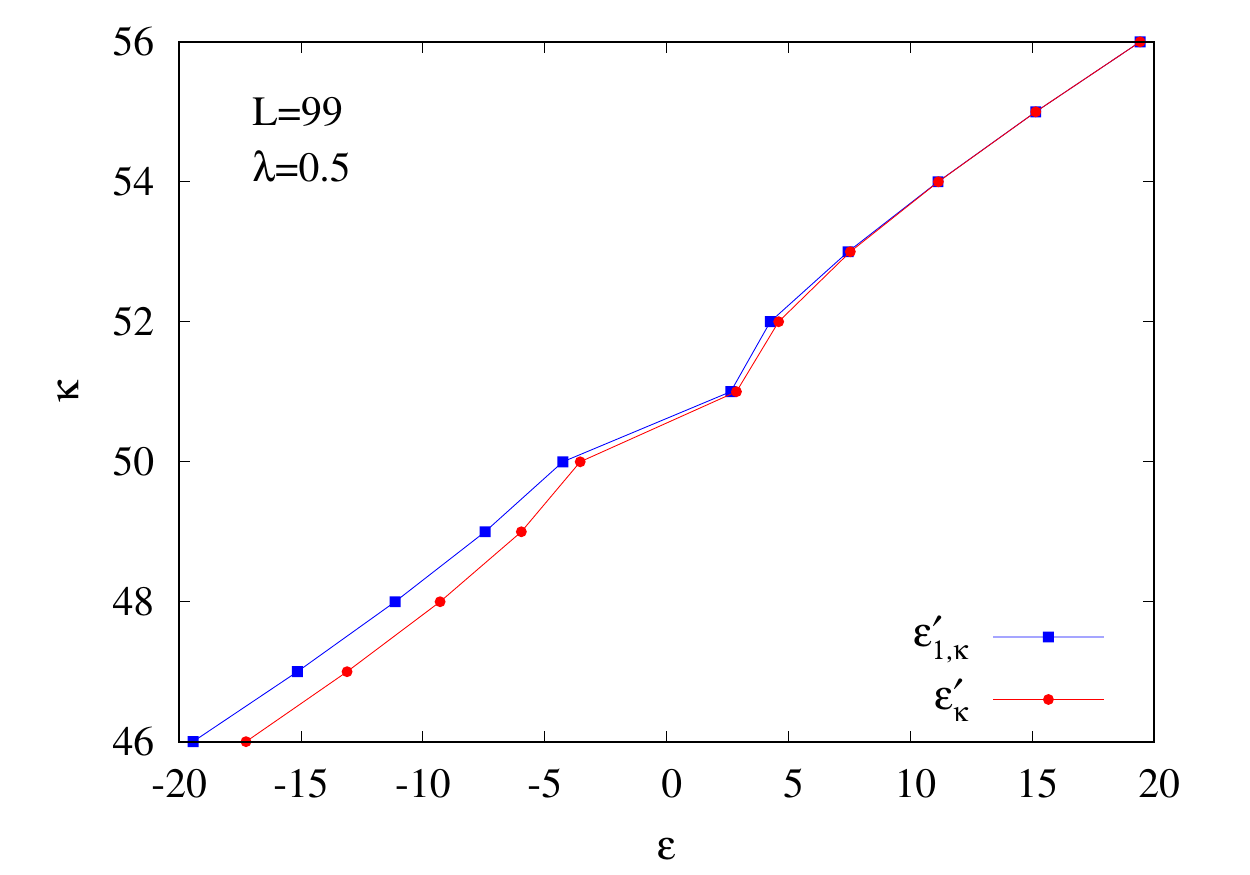}
\includegraphics[width=0.49\textwidth]{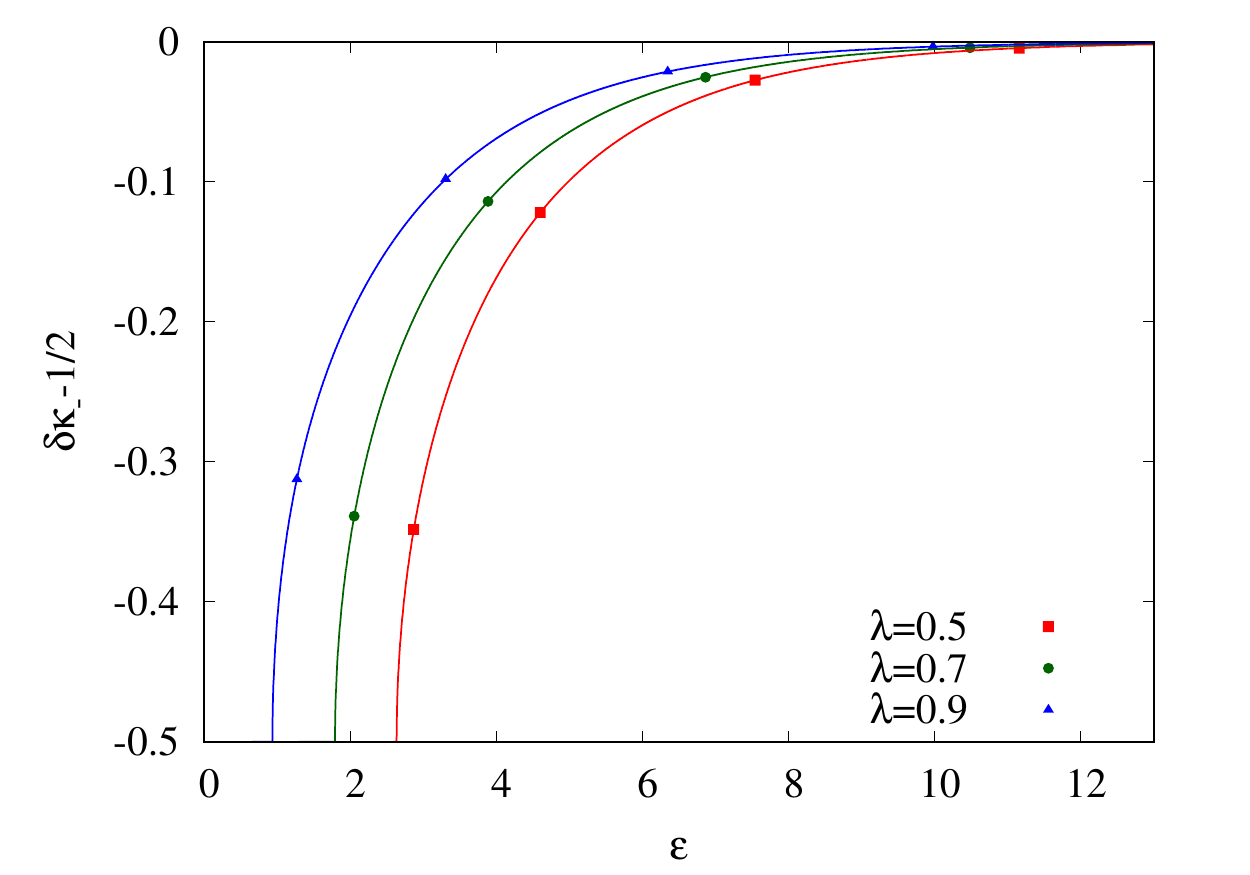}
\caption{Left: inverted single-particle entanglement spectra $\varepsilon'_{1,\kappa}$ and $\varepsilon'_{\kappa}$
for $\lambda=0.5$. Right: spectral shift for $\varepsilon>\varepsilon_+$ and various values of $\lambda$. The symbols show the
numerically obtained results (see text), while the solid lines correspond to the analytical formula \eqref{dkepspm}.
The top/bottom rows correspond to the even/odd cases wih $L=100$ and $L=99$, respectively.}
\label{fig:dkepsdef}
\end{figure}

Finally, we present our results for the spectral shift, obtained as the integral of the density \eqref{drhopm}.
It turns out that this can be evaluated in a closed form and yields
\eq{
\delta\kappa_\pm(\varepsilon) = \frac{1}{2\pi}
\left(\arctan \sqrt{\lambda^2 \cosh^2 \left(\frac{\varepsilon}{2}\right)-1} \mp
\arctan \frac{\sqrt{\lambda^2 \cosh^2 \left(\frac{\varepsilon}{2}\right)-1}}{\sqrt{1-\lambda^2}}
\right),
\label{dkepspm}}
where the integration constant has been chosen such that $\delta \kappa_\pm(\varepsilon_+)=0$. Note that for $\varepsilon\to\infty$ one has $\delta \kappa_+ \to 0$ and $\delta \kappa_- \to 1/2$.
The comparison to the numerical data is carried out in a similar fashion as in the homogeneous case. First, one inverts the mixed- and pure-state spectra, $\varepsilon'_\kappa$ and $\varepsilon'_{1,\kappa}$, as shown in the left panel of Fig. \ref{fig:dkepsdef} for both even/odd cases. In order to subtract the blue curve from the red one, we need again an interpolation of the data at the proper $\varepsilon$ values. This, however, is simply obtained by combining the homogeneous ansatz \eqref{k1eps} with the eigenvalue relation \eqref{epsdef}. The difference of the counting functions obtained this way is shown by the symbols in the right panel of Fig. \ref{fig:dkepsdef}, for the positive part ($\varepsilon>\varepsilon_+$) of the spectrum. Note that in the odd case we have to subtract the constant $1/2$ from $\delta\kappa_-$, to correctly reproduce the asymptotics shown by the data. In contrast, the negative part of the spectra is properly described by the functions $\delta\kappa_+ -1/2$ and $\delta \kappa_-$, respectively. The constant shifts between the positive/negative parts are actually due to the delta functions in \eqref{drhoeo}.

%%%%%%%%%%%%%%%%%%%%%%%%%%%%%
%%%%%%%%%%%%%%%%%%%%%%%%%%%%%

\section{Discussion\label{sec:discussion}}

We studied the variation of the ground-state entropy induced by the presence of a zero mode in free-fermion chains with a conformal defect. The underlying change in the density of the entanglement spectrum is calculated analytically via the FCS and the related resolvent function.
In the homogeneous case, we reobtained the result of Ref. \cite{KVW17} for the zero-mode entanglement entropy in an alternative way, and generalized it to the R\'enyi entropies. The calculations can also be extended to the defect, by making use of the relation that connects the spectrum to that of the homogeneous chain. We find excellent agreement between the analytical and numerical results.

A particular feature observed for the defect is the presence of parity effects in the zero-mode entropy, such that $\delta S_e$ and $\delta S_o$ differs for even/odd half-chain sizes. In fact, 
a closer inspection reveals that the parity effects are present in the mixed-state entropy $S$, while the pure-state entropies $S_0$
and $S_1$ show only alternations that vanish as $1/L$. Remarkably, finite parity effects were observed also in the pure ground state of a simple hopping defect \cite{STHS22}, which does not support a zero mode. Moreover, the qualitative behaviour of the parity term as a function of the transmission amplitude seems rather similar to our result $\delta S_e-\delta S_o$, shown in the inset of Fig.~\ref{fig:dSdef}. However, a quick numerical comparison of the two cases reveals, that the two functions are slightly different.
Indeed, the parity effects for the hopping defect must originate from the fact, that the relation analogous to \eqref{epsdef} is satisfied only approximately there. It would be interesting to see, whether the methods employed here could be generalized to understand this case.

One should remark that the parity effects found for the conformal defect can be directly generalized to the continuum setting, i.e. for a junction of quantum wires described by a scale-invariant scattering matrix. Indeed, the main relation \eqref{epsdef} connecting the defect spectrum to the homogeneous one remains unchanged \cite{CMV12}. Furthermore, as shown in Appendix \ref{app:continuum} by a direct calculation, while for even particle numbers the entanglement spectra are particle-hole symmetric, for odd occupations the symmetry is explicitly broken for the defect. This is exactly the same mechanism as the one observed for the lattice problem in Sec. \ref{sec:Defect}.

It would also be nice to extend the results for the conformal defect to arbitrary subsystem ratios. Again, the major bottleneck is to find the generalization of \eqref{epsdef}, to relate the defect problem to the homogeneous one.
%In fact, in this general setting the first step would be to understand the behaviour of the entropy without the zero mode.
Finally, it would be important to check the universality of the results by investigating more complicated defect problems supporting a zero mode. We believe that our results could be reproduced by a pure boundary CFT approach, similar to the one employed in Appendix \ref{app:CFT}, and it would be a fingerprint of universality.

Another interesting direction to explore is the evolution of entanglement across the defect, starting from a mixed initial state due to a zero-mode degeneracy. Indeed, it contains information both about the spreading of quantum correlations as well as the classical ones due to the mixture. The methods developed here could be combined with recent results on the time evolution of correlations \cite{gds-22,gds-23} and FCS across a defect \cite{glc-20,GZI22}, and used to analyse different quench protocols, generalizing the results of Refs. \cite{EP12,GE20,ce-22,csrc-23,wwr-18}.

\begin{acknowledgments}
VE acknowledges funding from the Austrian Science Fund (FWF) through Project No. P35434-N. LC acknowledges support from ERC under Consolidator grant number 771536 (NEMO).
\end{acknowledgments}

\appendix

\section{CFT derivation of the FCS ratio\label{app:CFT}}

In this appendix, we provide a proof of the formula Eq.~\eqref{detratio} employing CFT techniques. We first give a field theoretic characterization of our system in the absence of a defect, and then we compare the FCS of the vacuum and the one-particle state.
We consider the CFT of the Dirac fermions (see Ref. \cite{DiFrancesco-97}) on a spatial box of length $L$. We denote by $\ket{0}$ its vacuum, that we represent as a two-dimensional strip geometry parametrized by a complex variable $z$
\be
\text{Re}(z) \in [0,L].
\ee
The boundary conditions at the edges are chosen of the Dirichlet type (the density of fermions vanishes at that points), and we depict them as boundary lines $\text{Re}(z) = 0,L$ extended over the euclidean time (for simplicity we use $L$ for the chain size, in contrast to $2L$ used in the main text). We consider the subregion
\be
A = [0,\ell],
\ee
that is an interval attached to the boundary, and we study its FCS. For instance, given the $U(1)$ symmetry corresponding to the imbalance of particle and antiparticles, we construct its restriction over $A$ as
\be
\hat{N}_A = \int^\ell_0 dx \ \Psi^\dagger(x)\Psi(x),
\ee
with $\Psi(x)$ being the Dirac field. The quantum fluctuations of $\hat{N}_A$ in the vacuum state, are encoded in the full counting statistics
\be
\bra{0} e^{i\alpha \hat{N}_A}\ket{0}.
\ee

As shown in \cite{gnt-04}, a clever way to compute the expectation value above is via bosonization techniques. For instance, the following correspondence holds
\be
e^{i\alpha \hat{N}_A} \sim V_{\alpha/2\pi}(z=0)V_{-\alpha/2\pi}(z=\ell), \quad \alpha \in [-\pi,\pi],
\ee
with $V_{\pm \alpha/2\pi}(z,\bar{z})$ being vertex operators, whose expression in terms of the chiral modes $\phi,\bar{\phi}$ is
\be
V_{\pm \alpha/2\pi}(z,\bar{z}) = e^{\pm i\alpha/2\pi(\phi(z)+\bar{\phi}(\bar{z}))}.
\ee
Due to the choice of the boundary condition at $z=0$, one can show that the (boundary) scaling dimension of $V_{\alpha/2\pi}(z=0)$ vanishes, and from now on we just discard its insertion. In contrast, the (bulk) scaling dimension of $V_{-\alpha/2\pi}(z=\ell)$ has a nontrivial value given by \cite{DiFrancesco-97}
\be
\Delta = \left(\frac{\alpha}{2\pi}\right)^2.
\ee
Finally, we get
\be
\bra{0} e^{i\alpha \hat{N}_A}\ket{0} \sim \bra{0} V_{-\alpha/2\pi}(z=\ell)\ket{0},
\ee
namely a one-point function of a scalar primary field in a strip geometry, that we aim to compute below.
%%%%
\begin{figure}[htb]
\center
\includegraphics[width=0.59\textwidth]{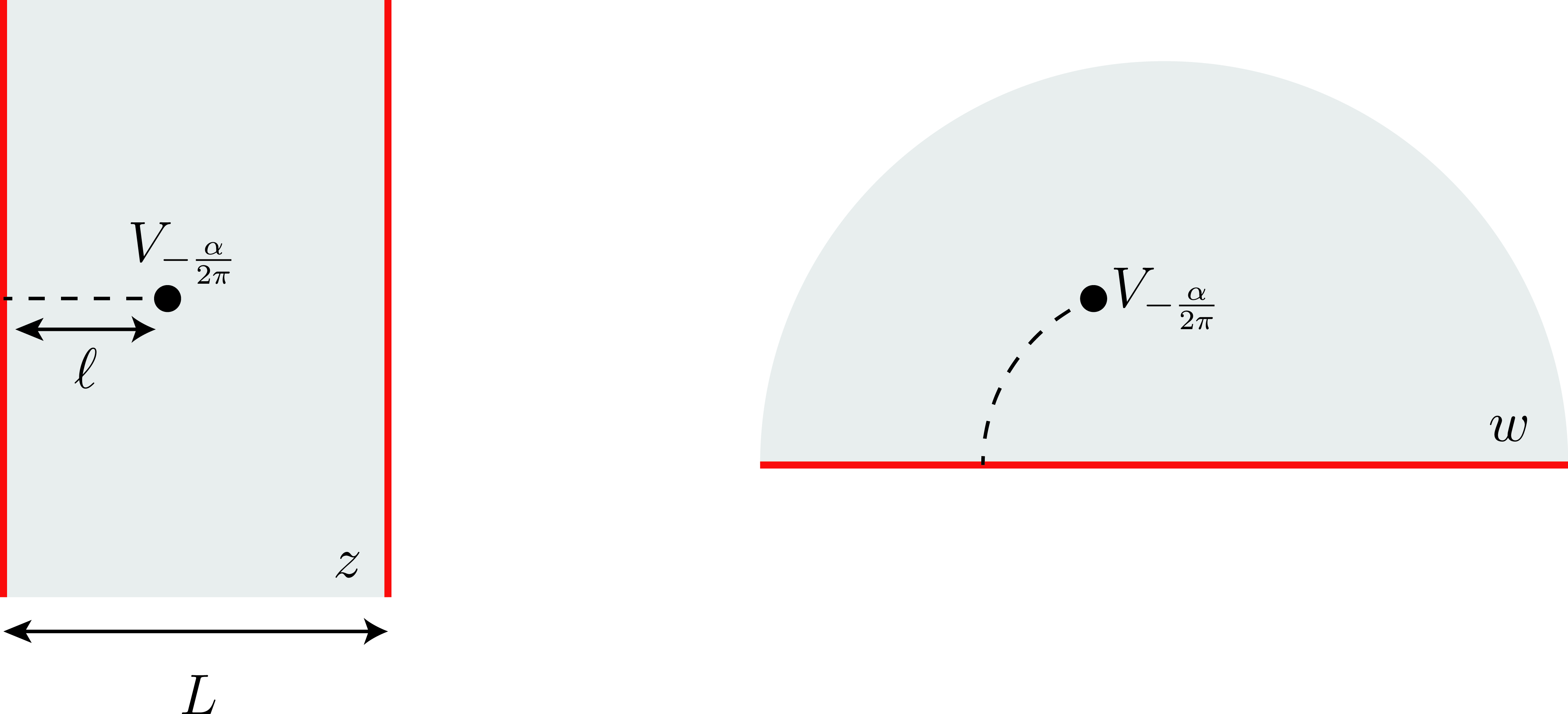}
\caption{One-point function of $V_{-\alpha/2\pi}$ in the strip geometry (left) and upper half-plane (right), computed in the vacuum state $\ket{0}$. The red lines at the edges of the strip represent the boundary points extended over the euclidean time, and they are mapped onto the real axis via $z\rightarrow w$.}
\label{fig:Hom_CFT}
\end{figure}
%%%%
To do so, we first employ the conformal transformation
\be
w = -e^{-i\frac{\pi}{L}z},
\ee
which maps the strip geometry onto the upper half-plane (UHP) $\text{Im}(w)\geq 0$. In this new geometry, the one-point function is fixed by symmetries\footnote{The upper half-plane is invariant under $w\rightarrow w+\epsilon$ (real translation) and $w\rightarrow e^{\epsilon}w$ (scaling), $\epsilon \in \mathbb{R}$} and it is given by
\be
\la V_{-\alpha/2\pi}(w,\bar{w})\ra_{\text{UHP}} = \frac{1}{(w-\bar{w})^{\Delta}}.
\ee
We summarize the construction above in Fig. \ref{fig:Hom_CFT}, which gives a pictorial representation of the two geometries considered above.

To proceed further, we should go back to the initial geometry, and, using the transformation law of primary operators, we express
\be
\bra{0}V_{-\alpha/2\pi}(z,\bar{z})\ket{0} = \left| \frac{dw}{dz} \right|^{\Delta} \la V_{-\alpha/2\pi}(w,\bar{w})\ra_{\text{UHP}} = \left( \frac{\pi}{2L} \frac{1}{\sin \frac{\pi(z+\bar{z})}{2L}}\right)^{\Delta}.
\ee
Putting everything together, we get for the full-counting statistics of the vacuum as
\be
\bra{0} e^{i\alpha \hat{N}_A}\ket{0} \sim \left( \frac{\pi}{2L} \frac{1}{\sin \frac{\pi \ell}{L}}\right)^{\left( \frac{\alpha}{2\pi}\right)^2},
\ee
which holds up to a non-universal ($\alpha$-dependent) proportionality constant.

\begin{figure}[htb]
\center
\includegraphics[width=0.45\textwidth]{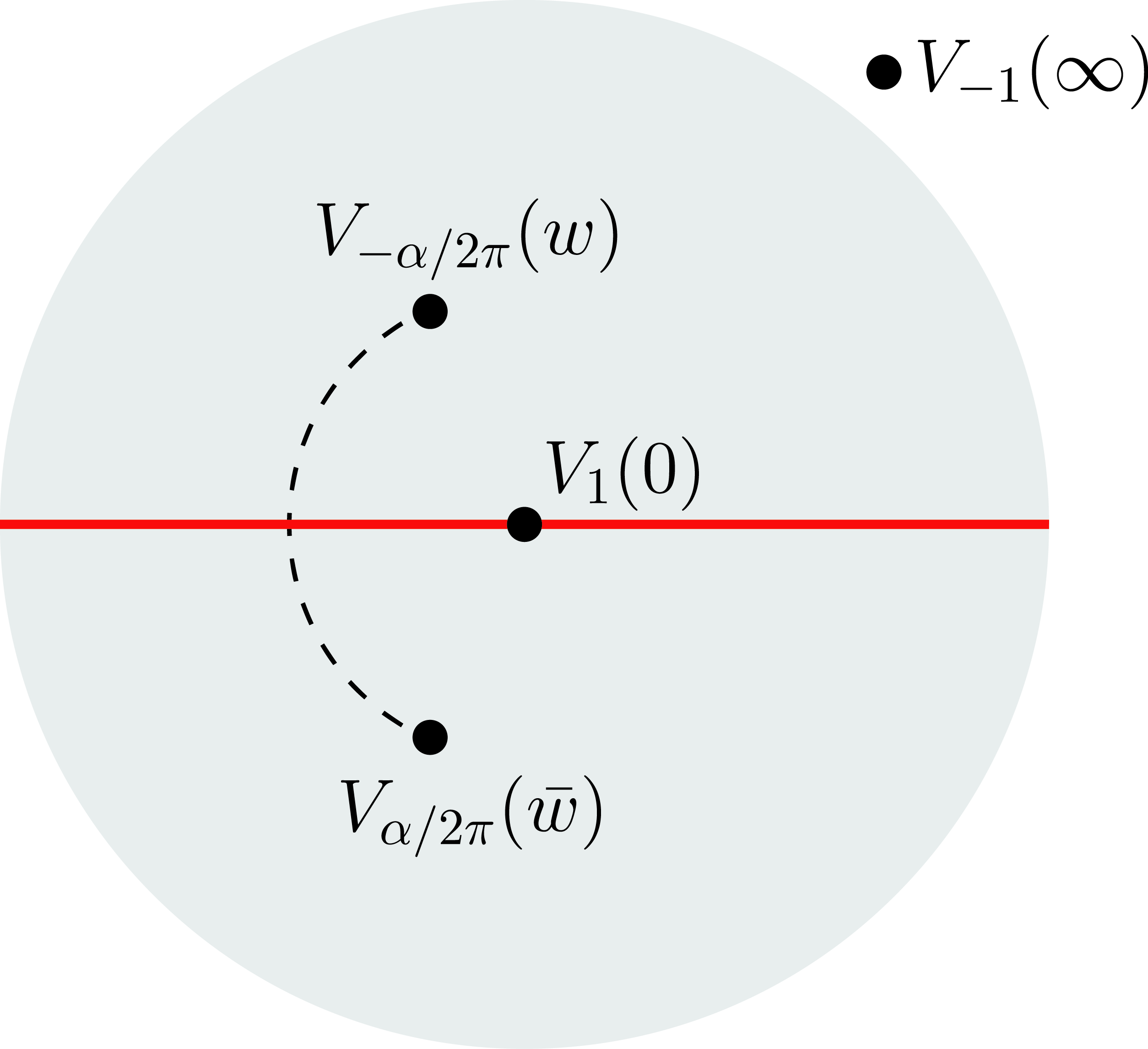}
\caption{Chiral CFT on the plane, corresponding to the full CFT on the UHP via unfolding. $V_1(0)$ and $V_{-1}(\infty)$ represent the ket $\ket{1}$ and bra $\bra{1}$ associated to the one-particle excited state. The insertion of the chiral vertex operators $V_{-\alpha/2\pi}(w),V_{\alpha/2\pi}(\bar{w})$ is related to the scalar vertex operator $V_{-\alpha/2\pi}(w,\bar{w})$ of the UHP.}
\label{fig:Exc_Hom}
\end{figure}

We now repeat the same calculation for the excited state made of a single particle just above the Fermi sea (the vacuum), denoted here by $\ket{1}$ (see also Ref. \cite{acb-11}). A powerful approach to tackle this problem relies on the equivalence between the CFT in the UHP, to its chiral counterpart on the whole complex plane (see Refs. \cite{DiFrancesco-97,Cardy-04} for details), a procedure called \textit{unfolding}. In this way, one can employ radial quantization, describing the excited state via an insertion of a local field at $w=0,\infty$ on the conformal vacuum in planar geometry. For our purpose, we need to insert the chiral vertex operators $V_{\pm 1}$, with conformal dimension $1$, at $w=0,\infty$, corresponding to the bra/ket of the one-particle state (as explained in \cite{acb-11}). Through the unfolding procedure, the antiholomorphic fields inserted in the upper half-plane are mapped onto holomorphic fields at their specular position wrt the real axis. In particular, this leads to the replacement
\be
V_{-\alpha/2\pi}(w,\bar{w}) \rightarrow V_{-\alpha/2\pi}(w)V_{\alpha/2\pi}(\bar{w}) 
\ee
inside the correlation functions. In the end, we need the 4-point function
\be
\la V_{-1}(\infty)V_{\alpha/2\pi}(\bar{w}) V_{-\alpha/2\pi}(w)V_{1}(0) \ra
\ee
evaluated in the planar geometry, and the result is \cite{DiFrancesco-97,crc-20}
\be
\frac{\la V_{-1}(\infty)V_{\alpha/2\pi}(\bar{w}) V_{-\alpha/2\pi}(w)V_{1}(0) \ra}{\la V_{-1}(\infty)V_{1}(0) \ra} = \la V_{\alpha/2\pi}(\bar{w}) V_{-\alpha/2\pi}(w) \ra \times \left( \frac{\bar{w}}{w}\right)^{\alpha/2\pi}.
\ee
The previous expression gives the expectation value of $V_{-\alpha/2\pi}(w,\bar{w})$ in the UHP, and the denominator $\la V_{-1}(\infty)V_{1}(0)\ra$ ensures the proper normalization of the state $\ket{1}$.
We represent the unfolded geometry and the field insertions in Fig. \ref{fig:Exc_Hom}, which summarizes this construction.
%%%%
%%%%

The last step is the map $w\rightarrow z$, which brings back to the initial geometry, and we get
\begin{align}
\bra{1}V_{-\alpha/2\pi}(z,\bar{z})\ket{1} = &\left| \frac{dw}{dz}\right|^{\Delta} \frac{\la V_{-1}(\infty)V_{\alpha/2\pi}(\bar{w}) V_{-\alpha/2\pi}(w)V_{1}(0) \ra}{\la V_{-1}(\infty)V_{1}(0) \ra} =\\
&\bra{0}V_{-\alpha/2\pi}(z,\bar{z})\ket{0} \times e^{i\frac{\alpha}{2L}(z+\bar{z})}.
\end{align}
In the end, we express the ratio of FCS as
\be\label{eq:FCS_ratio}
\frac{\bra{1}e^{i\alpha \hat{N}_A}\ket{1}}{\bra{0}e^{i\alpha \hat{N}_A}\ket{0}} = e^{i\alpha \frac{\ell}{L}},
\ee
which is the main result employed in Section \ref{sec:hom}, see Eq. \eqref{detratio}. From our prediction, one learns that the difference of connected moments between the two states is universal. For instance one has
\be
\bra{1}\hat{N}_A\ket{1} - \bra{0}\hat{N}_A\ket{0} = \frac{\ell}{L},
\ee
while the difference of the other moments is vanishing. We finally mention that the same result was already obtained for the ring geometry (periodic boundary conditions) in \cite{crc-20}, with similar techniques. The origin of this match, which is not obvious a priori, is ultimately found in the equivalence between the CFT on the UHP and its chiral counterpart on the plane. 

To conclude this appendix, we point out a technical, albeit fundamental, observation. The CFT calculation presented here refers to $\alpha\in [-\pi,\pi]$, while other real values of $\alpha$ can be obtained via the periodic property $\alpha \rightarrow \alpha+2\pi$, which comes from the definition. However, in the main text we have implicitly analytically continued the result over complex values of $\alpha$, as the change of variable $z=\frac{1}{1-e^{i\alpha}}$ in Eq. \eqref{z} was employed for real $z$. We conjecture that this procedure is justified. Nevertheless, we point out that while the analytical continuation of Eq. \eqref{eq:FCS_ratio} over the whole complex plane is clearly possible, it does not coincide with the actual value of the ratio of FCS, as it does not satisfy the symmetry under $\alpha \rightarrow \alpha+2\pi$. Moreover, we argue that the match between the two is present only in the strip $\text{Re}(\alpha) \in [-\pi,\pi]$. A rigorous analysis of the analytic properties of the FCS is nevertheless beyond the purpose of this work, and we refer to Ref. \cite{fg-20} for a similar discussion.

\section{Relation between entanglement spectra \label{app:epsdef}}

In this appendix we prove the relation \eqref{epsdef}.
Let us consider a Fermi sea ground state with the lowest $N$ modes occupied and the rest empty. The correlation matrix for the defect reads
\eq{
C'_{mn} = \sum_{k=1}^{N} \phi'_k(m)\phi'_k(n) \, ,
}
with the eigenvectors given in \eqref{phikdef}.
We shall focus on a half-chain partition with $A=\left[1,L\right]$. The key step is to consider the product $C'_A(1-C'_A)$, with matrix elements given by
\eq{
\left[C'_A(1-C'_A)\right]_{mn}=
\sum_{k=1}^{N}\sum_{l=N+1}^{2L} \alpha^2_k \alpha^2_l \,
\phi_k(m) A_{kl} \, \phi_l(n) \, ,
\label{C1-C}}
where we introduced the overlap matrix
\eq{
A_{kl}=\sum_{j=1}^{L} \phi_k(j) \phi_l(j) \, .
\label{Akl}}

Now the main observation is that \eqref{C1-C} depends on the defect only via the factor
\eq{
\alpha^2_k \alpha^2_l=
\begin{cases}
(1+\sqrt{1-\lambda^2})^2& \textrm{$k-l$ even} \\
\lambda^2 & \textrm{$k-l$ odd}
\end{cases},
\label{akal}}
where we used \eqref{ab}. Furthermore, we can also show that the overlap matrix has a checkerboard structure.
Inserting the explicit form \eqref{phihom} of the eigenvectors, the sum \eqref{Akl} can be carried out as
\eq{
A_{kl}= \frac{\sin \left[\frac{\pi}{2}(m-n)\right]}{4L\sin\left[\frac{\pi}{4L}(m-n)\right]}
-\frac{\sin \left[\frac{\pi}{2}(m+n)\right]}{4L\sin\left[\frac{\pi}{4L}(m+n)\right]} \, .
}
Thus one can immediately see, that the matrix elements $A_{kl}$ are nonvanishing only for $k-l$ odd, and together with \eqref{akal} this leads to the relation
\eq{
C'_A(1-C'_A)=
\lambda^2 \, C_A(1-C_A) \, .
\label{eq:Rel_Spec}}
Rewriting in terms of the eigenvalues one has
\eq{
\zeta'_{\kappa}(1-\zeta'_{\kappa})=
\lambda^2 \, \zeta_{\kappa}(1-\zeta_{\kappa}) \, .
\label{zetarel}}
Finally, using \eqref{epszeta} one obtains
\eq{
\zeta_{\kappa}(1-\zeta_{\kappa})=
\frac{1}{4\cosh^2\frac{\varepsilon_\kappa}{2}}
}
and similarly for the defect eigenvalues. Inserting into \eqref{zetarel}, one arrives at the relation \eqref{epsdef} reported in the main text.

\section{Even/odd effects for the Schrödinger junction\label{app:continuum}}

Here, we consider a Fermi gas in a finite geometry with a conformal defect in the middle, which is dubbed as Schrödinger junction \cite{cmv-11,cmv-11a,CMV12}. This system is closely related to the chain \eqref{H}, and one expects that the universal features of the two models are captured by the same field theory (see Ref. \cite{cmc-22}). For instance, the entropy of the state with the first $N$ levels filled is known to diverge logarithmically in $N$, and the prefactor is the same for the CFT \cite{cmc-22} and the chain \cite{ep-10}. In this appendix, we show the presence of peculiar even/odd effects as $N$ is varied. The mechanism we find is equivalent to the one of the chain in Sec. \ref{sec:Defect}, and, as we show, it gives rise to the same universal features.

The points of the Schrödinger junction are parametrized by a pair
\be
(x,j), \quad x \in [0,L], \quad  j=1,2,
\ee
where $j$ labels the two wires and $x$ the associated spatial position. The bulk Hamiltonian, as a function of the fermionic field $\Psi_j(x)$, is 
\be\label{eq:SchJun_Ham}
H = \sum^{2}_{j=1} \int^L_{0} dx\frac{1}{2}\left( \partial_x \Psi^\dagger_j(x)\right)\left( \partial_x \Psi_j(x)\right),
\ee
and the boundary conditions have to be specified at $x=0,L$. At $x=0$ we consider a scale invariant scattering matrix
\be\label{eq:Smatrix}
S = \begin{pmatrix} \sqrt{1-\lambda^2} & \lambda \\ \lambda & -\sqrt{1-\lambda^2}\end{pmatrix}, \quad \lambda \in [0,1]
\ee
which couples the two wires explicitly, and it corresponds to the conformal defect. Here, $\lambda$ is the transmission amplitude and its physical meaning is the same as for the chain of Sec. \ref{sec:Defect}. At $x=L$, we choose Dirichlet boundary conditions, namely
\be
\Psi_j(L)=0.
\ee

We now aim to characterize the eigenstates of $H$. To do so, we have to find first its single-particle levels, and then specify their occupation numbers. A convenient strategy is a change of basis which diagonalizes $S$, whose eigenvalues are $\pm 1$, via a unitary $2 \times 2$ matrix $\mathcal{U}$. This amounts to the introduction of a pair of unphysical fields $\varphi_1(x),\varphi_2(x)$ as
\be\label{eq:unphys}
\Psi_i(x)= \sum_{j=1}^2\mathcal{U}_{ij}\varphi_j(x),
\ee
which are decoupled and satisfy Neumann(N)/Dirichlet(D) boundary conditions at $x=0$ respectively
\be
\partial_x\varphi_1(0) = 0, \quad \varphi_2(0)=0.
\ee
For these two boundary conditions, we denote the single-particle eigenfunctions as
\be
\phi^{N}_n(x) = \sqrt{\frac{2}{L}}\cos \frac{\left( n-1/2\right)\pi x}{L}, \quad
\phi^{D}_n(x) = \sqrt{\frac{2}{L}}\sin \frac{n\pi x}{L}, \quad n=1,2,\dots
\ee

We now consider a Fermi sea made by 
 the first $N_{N/D}$ levels filled with boundary conditions $N/D$ respectively, and the total particle number is $N = N_{N}+N_{D}$. The correlation function of the unphysical fields is thus
\be
\la \varphi^\dagger_j(x)\varphi_{j'}(x')\ra = \delta_{jj'} \times \begin{cases} C_{N}(x,x'), \quad j=1 \\C_{D}(x,x'), \quad j=2, \end{cases}
\ee
 where
 \be
C_{N}(x,x') =  \sum^{N_N}_{n=1} \phi^{N}_n(x)\phi^{N}_n(x'), \quad C_{D}(x,x') =  \sum^{N_D}_{n=1} \phi^{D}_n(x)\phi^{D}_n(x')
\ee
Now, for a fixed particle number $N$, we focus on the lowest energy state, which corresponds to $N_N=N_D=N/2$ for $N$ even and $N_N=N_D+1=(N+1)/2$ for $N$ odd. Going back to the physical fields $\Psi_j$, one can eventually express the correlation function as  (see Ref. \cite{cmc-22} for details)
\be
C'_{jj'}(x,x') \equiv \la \Psi_j^\dagger(x)\Psi_{j'}(x') \ra = \left( \frac{1+S}{2}\right)_{jj'} C_{N}(x,x') + \left(\frac{1-S}{2}\right)_{jj'}C_{D}(x,x').
\label{eq_Kernel}
\ee

We construct the restricted kernels associated to the first and the second wire, denoted here by $C'_{11}(x,x')$ and $C'_{22}(x,x')$, and compute their spectrum. Following \cite{cmc-22}, we consider a $N$ dimensional subspace of $L^2([0,L])$ spanned by the set of wave functions $\{\phi^N_n(x)\}^{N_N}_{n=1} \cup \{\phi^D_n(x)\}^{N_D}_{n=1}$, that we take as a (non-orthonormal) basis. For instance, one can show that the kernel $C'_{jj'}(x,x')$ acts non-trivially in the subspace considered,  while it vanishes on the orthogonal subspace. Thus, by projecting the kernel on this subspace, we can access directly its non-vanishing spectrum.
We do so, and we end up with the following (block) matrix representation%
\be
C_{N} \simeq \begin{pmatrix} 1 & Q\\ 0 & 0\end{pmatrix}, \quad C_D \simeq \begin{pmatrix} 0 & 0\\ Q^\dagger & 1\end{pmatrix},
\ee
with $Q_{nn'}$ a $N_N\times N_D$ rectangular matrix defined as
\be
Q_{nn'} \equiv \int^L_0dx \ \phi^N_n(x)\phi^D_{n'}(x), \quad n=1,\dots,N_N, \quad n' = 1,\dots,N_D,
\ee
corresponding to the scalar product between the non-orthogonal basis elements. Here, the symbol $\simeq$ refers to the equivalence of the non-zero spectrum, i.e. the set of non-zero eigenvalues. In this way, the restricted kernel of the first wire is expressed as a $N\times N$ matrix
\be
C'_{11} \simeq \frac{1+\sqrt{1-\lambda^2}}{2} \begin{pmatrix} 1 & Q\\ 0 & 0\end{pmatrix} + \frac{1-\sqrt{1-\lambda^2}}{2}\begin{pmatrix} 0 & 0\\ Q^\dagger & 1\end{pmatrix}.
\ee

At this point, a simple computation shows that the spectra of $C'_{11}$ at any $\lambda$ is related to the one of $C_{11}$, obtained for $\lambda=1$, as \cite{cmc-22}
\be\label{eq:Rel_Spec_Sch}
C'_{11}(1-C'_{11})= \lambda^2 C_{11}(1-C_{11}),
\ee
which is equivalent to Eq. \eqref{eq:Rel_Spec}. Since the relation \eqref{eq:Rel_Spec_Sch} is not invertible, the spectrum of $C_{11}$ does not fix unambigously the one of $C'_{11}$. In particular, as we will show below, $C'_{11}$ and $1-C'_{11}$ have different spectra for $N$ odd, and the (particle-hole) symmetry $C'_{11} \leftrightarrow 1-C'_{11}$ is explicitly broken. For the sake of convenience, we introduce the $N\times N$ matrix
\be
\Gamma = 1-2C'_{11},
\ee
so that the particle-hole symmetry corresponds to $\Gamma \leftrightarrow -\Gamma$. To compute its spectrum, we
express the characteristic polynomial of $\Gamma$
\be
\text{det}\left( y-\Gamma\right) = \text{det} \begin{pmatrix}
y+\sqrt{1-\lambda^2} & (1+\sqrt{1-\lambda^2})Q\\
(1-\sqrt{1-\lambda^2})Q^\dagger & y-\sqrt{1-\lambda^2}
\end{pmatrix},
\label{charpol}
\ee
that is the determinant of a block matrix, and it can be evaluated thanks to the property
\be
\text{det}\begin{pmatrix}
A & B \\ C & D
\end{pmatrix} = \text{det}(A)\text{det}(D-CA^{-1}B).
\ee
When $N$ is even, every block in Eq. \eqref{charpol} is $N/2 \times N/2$ and the characteristic polynomial is
\be
\text{det}\left( y-\Gamma\right) = \text{det}\left( y^2-1+\lambda^2(1-Q^\dagger Q)\right).
\ee
Since it is symmetric under $y\rightarrow -y$, it means that the particle hole symmetry $\Gamma \leftrightarrow -\Gamma$ is preserved. In contrast, when $N$ is odd, since $N_N = N_D+1$ we get
\be
\text{det}\left( y-\Gamma\right) = (y+\sqrt{1-\lambda^2})\text{det}\left( y^2-1+\lambda^2(1-Q^\dagger Q)\right).
\ee
In other words, there is a single eigenvalue $-\sqrt{1-\lambda^2}$ of $\Gamma$, corresponding to the eigenvalue $\frac{1+\sqrt{1-\lambda^2}}{2}$ of $C'_{11}$, which breaks explicitly the particle-hole symmetry, while the other ones preserve it. Similar calculation can be performed also for the second wire, and one gets an eigenvalue $\frac{1-\sqrt{1-\lambda^2}}{2}$ for $C'_{22}$ whenever $N$ is odd, which implies an asymmetry between the two wires. Its origin can be easily traced back to the choice of the scattering matrix \eqref{eq:Smatrix} which is asymmetric under the exchange of the wires. In summary, we find precisely the same mechanism in the continuum as for the lattice model in Sec. \ref{sec:Defect}, which leads to the parity effects in the entropy.

%Indeed, given the mixture of the Fermi seas with $N$ and $N+1$ particles, corresponding to the zero-temperature state with a zero mode, its von Neumann entropy shows universal even/odd effects as $N$ is varied. We also conjecture that the proof presented in this appendix might be applied, with minor technical modifications, to the lattice, but this is beyond our scope.

\newpage

\bibliography{zeromodedef}

%apsrev4-2.bst 2019-01-14 (MD) hand-edited version of apsrev4-1.bst
%Control: key (0)
%Control: author (8) initials jnrlst
%Control: editor formatted (1) identically to author
%Control: production of article title (0) allowed
%Control: page (0) single
%Control: year (1) truncated
%Control: production of eprint (0) enabled
\begin{thebibliography}{73}%
\makeatletter
\providecommand \@ifxundefined [1]{%
 \@ifx{#1\undefined}
}%
\providecommand \@ifnum [1]{%
 \ifnum #1\expandafter \@firstoftwo
 \else \expandafter \@secondoftwo
 \fi
}%
\providecommand \@ifx [1]{%
 \ifx #1\expandafter \@firstoftwo
 \else \expandafter \@secondoftwo
 \fi
}%
\providecommand \natexlab [1]{#1}%
\providecommand \enquote  [1]{``#1''}%
\providecommand \bibnamefont  [1]{#1}%
\providecommand \bibfnamefont [1]{#1}%
\providecommand \citenamefont [1]{#1}%
\providecommand \href@noop [0]{\@secondoftwo}%
\providecommand \href [0]{\begingroup \@sanitize@url \@href}%
\providecommand \@href[1]{\@@startlink{#1}\@@href}%
\providecommand \@@href[1]{\endgroup#1\@@endlink}%
\providecommand \@sanitize@url [0]{\catcode `\\12\catcode `\$12\catcode
  `\&12\catcode `\#12\catcode `\^12\catcode `\_12\catcode `\%12\relax}%
\providecommand \@@startlink[1]{}%
\providecommand \@@endlink[0]{}%
\providecommand \url  [0]{\begingroup\@sanitize@url \@url }%
\providecommand \@url [1]{\endgroup\@href {#1}{\urlprefix }}%
\providecommand \urlprefix  [0]{URL }%
\providecommand \Eprint [0]{\href }%
\providecommand \doibase [0]{https://doi.org/}%
\providecommand \selectlanguage [0]{\@gobble}%
\providecommand \bibinfo  [0]{\@secondoftwo}%
\providecommand \bibfield  [0]{\@secondoftwo}%
\providecommand \translation [1]{[#1]}%
\providecommand \BibitemOpen [0]{}%
\providecommand \bibitemStop [0]{}%
\providecommand \bibitemNoStop [0]{.\EOS\space}%
\providecommand \EOS [0]{\spacefactor3000\relax}%
\providecommand \BibitemShut  [1]{\csname bibitem#1\endcsname}%
\let\auto@bib@innerbib\@empty
%</preamble>
\bibitem [{\citenamefont {Amico}\ \emph {et~al.}(2008)\citenamefont {Amico},
  \citenamefont {Fazio}, \citenamefont {Osterloh},\ and\ \citenamefont
  {Vedral}}]{afov-08}%
  \BibitemOpen
  \bibfield  {author} {\bibinfo {author} {\bibfnamefont {L.}~\bibnamefont
  {Amico}}, \bibinfo {author} {\bibfnamefont {R.}~\bibnamefont {Fazio}},
  \bibinfo {author} {\bibfnamefont {A.}~\bibnamefont {Osterloh}},\ and\
  \bibinfo {author} {\bibfnamefont {V.}~\bibnamefont {Vedral}},\ }\bibfield
  {title} {\bibinfo {title} {Entanglement in many-body systems},\ }\href
  {https://doi.org/https://doi.org/10.1103/RevModPhys.80.517} {\bibfield
  {journal} {\bibinfo  {journal} {Rev. Mod. Phys.}\ }\textbf {\bibinfo {volume}
  {80}},\ \bibinfo {pages} {517} (\bibinfo {year} {2008})}\BibitemShut
  {NoStop}%
\bibitem [{\citenamefont {Eisert}\ \emph {et~al.}(2010)\citenamefont {Eisert},
  \citenamefont {Cramer},\ and\ \citenamefont {Plenio}}]{ecp-10}%
  \BibitemOpen
  \bibfield  {author} {\bibinfo {author} {\bibfnamefont {J.}~\bibnamefont
  {Eisert}}, \bibinfo {author} {\bibfnamefont {M.}~\bibnamefont {Cramer}},\
  and\ \bibinfo {author} {\bibfnamefont {M.~B.}\ \bibnamefont {Plenio}},\
  }\bibfield  {title} {\bibinfo {title} {Colloquium: Area laws for the
  entanglement entropy},\ }\href {https://doi.org/10.1103/RevModPhys.82.277}
  {\bibfield  {journal} {\bibinfo  {journal} {Rev. Mod. Phys.}\ }\textbf
  {\bibinfo {volume} {82}},\ \bibinfo {pages} {277} (\bibinfo {year}
  {2010})}\BibitemShut {NoStop}%
\bibitem [{\citenamefont {Calabrese}\ \emph {et~al.}(2009)\citenamefont
  {Calabrese}, \citenamefont {Cardy},\ and\ \citenamefont {Doyon}}]{ccd-09}%
  \BibitemOpen
  \bibfield  {author} {\bibinfo {author} {\bibfnamefont {P.}~\bibnamefont
  {Calabrese}}, \bibinfo {author} {\bibfnamefont {J.}~\bibnamefont {Cardy}},\
  and\ \bibinfo {author} {\bibfnamefont {B.}~\bibnamefont {Doyon}},\ }\bibfield
   {title} {\bibinfo {title} {Entanglement entropy in extended quantum
  systems},\ }\href {https://doi.org/10.1088/1751-8121/42/50/500301} {\bibfield
   {journal} {\bibinfo  {journal} {J. Phys. A: Math. Theor.}\ }\textbf
  {\bibinfo {volume} {42}},\ \bibinfo {pages} {500301} (\bibinfo {year}
  {2009})}\BibitemShut {NoStop}%
\bibitem [{\citenamefont {Laflorencie}(2016)}]{Laflorencie-16}%
  \BibitemOpen
  \bibfield  {author} {\bibinfo {author} {\bibfnamefont {N.}~\bibnamefont
  {Laflorencie}},\ }\bibfield  {title} {\bibinfo {title} {Quantum entanglement
  in condensed matter systems},\ }\href
  {https://doi.org/10.1016/j.physrep.2016.06.008} {\bibfield  {journal}
  {\bibinfo  {journal} {Phys. Rep.}\ }\textbf {\bibinfo {volume} {646}},\
  \bibinfo {pages} {1} (\bibinfo {year} {2016})}\BibitemShut {NoStop}%
\bibitem [{\citenamefont {Calabrese}\ and\ \citenamefont
  {Cardy}(2009)}]{cc-09}%
  \BibitemOpen
  \bibfield  {author} {\bibinfo {author} {\bibfnamefont {P.}~\bibnamefont
  {Calabrese}}\ and\ \bibinfo {author} {\bibfnamefont {J.}~\bibnamefont
  {Cardy}},\ }\bibfield  {title} {\bibinfo {title} {Entanglement entropy and
  conformal field theory},\ }\href
  {https://doi.org/10.1088/1751-8113/42/50/504005} {\bibfield  {journal}
  {\bibinfo  {journal} {J. Phys. A Math. Theor.}\ }\textbf {\bibinfo {volume}
  {42}},\ \bibinfo {pages} {504005} (\bibinfo {year} {2009})}\BibitemShut
  {NoStop}%
\bibitem [{\citenamefont {Holzhey}\ \emph {et~al.}(1994)\citenamefont
  {Holzhey}, \citenamefont {Larsen},\ and\ \citenamefont {Wilczek}}]{hlw-94}%
  \BibitemOpen
  \bibfield  {author} {\bibinfo {author} {\bibfnamefont {C.}~\bibnamefont
  {Holzhey}}, \bibinfo {author} {\bibfnamefont {F.}~\bibnamefont {Larsen}},\
  and\ \bibinfo {author} {\bibfnamefont {F.}~\bibnamefont {Wilczek}},\
  }\bibfield  {title} {\bibinfo {title} {Geometric and renormalized entropy in
  conformal field theory},\ }\href
  {https://doi.org/10.1016/0550-3213(94)90402-2} {\bibfield  {journal}
  {\bibinfo  {journal} {Nucl. Phys. B}\ }\textbf {\bibinfo {volume} {424}},\
  \bibinfo {pages} {443} (\bibinfo {year} {1994})}\BibitemShut {NoStop}%
\bibitem [{\citenamefont {Calabrese}\ and\ \citenamefont
  {Cardy}(2004)}]{cc-04}%
  \BibitemOpen
  \bibfield  {author} {\bibinfo {author} {\bibfnamefont {P.}~\bibnamefont
  {Calabrese}}\ and\ \bibinfo {author} {\bibfnamefont {J.}~\bibnamefont
  {Cardy}},\ }\bibfield  {title} {\bibinfo {title} {Entanglement entropy and
  quantum field theory},\ }\href
  {https://doi.org/10.1088/1742-5468/2004/06/P06002} {\bibfield  {journal}
  {\bibinfo  {journal} {J. Stat. Mech.: Theory Exp.}\ }\textbf {\bibinfo
  {volume} {2004}}\bibinfo  {number} { (06)},\ \bibinfo {pages}
  {P06002}}\BibitemShut {NoStop}%
\bibitem [{\citenamefont {Vidal}\ \emph {et~al.}(2003)\citenamefont {Vidal},
  \citenamefont {Latorre}, \citenamefont {Rico},\ and\ \citenamefont
  {Kitaev}}]{vlrk-03}%
  \BibitemOpen
\bibfield  {number} {  }\bibfield  {author} {\bibinfo {author} {\bibfnamefont
  {G.}~\bibnamefont {Vidal}}, \bibinfo {author} {\bibfnamefont {J.~I.}\
  \bibnamefont {Latorre}}, \bibinfo {author} {\bibfnamefont {E.}~\bibnamefont
  {Rico}},\ and\ \bibinfo {author} {\bibfnamefont {A.}~\bibnamefont {Kitaev}},\
  }\bibfield  {title} {\bibinfo {title} {Entanglement in quantum critical
  phenomena},\ }\href {https://doi.org/10.1103/PhysRevLett.90.227902}
  {\bibfield  {journal} {\bibinfo  {journal} {Phys. Rev. Lett.}\ }\textbf
  {\bibinfo {volume} {90}},\ \bibinfo {pages} {227902} (\bibinfo {year}
  {2003})}\BibitemShut {NoStop}%
\bibitem [{\citenamefont {Affleck}\ \emph {et~al.}(2009)\citenamefont
  {Affleck}, \citenamefont {Laflorencie},\ and\ \citenamefont
  {S\o{}rensen}}]{ALS09}%
  \BibitemOpen
  \bibfield  {author} {\bibinfo {author} {\bibfnamefont {I.}~\bibnamefont
  {Affleck}}, \bibinfo {author} {\bibfnamefont {N.}~\bibnamefont
  {Laflorencie}},\ and\ \bibinfo {author} {\bibfnamefont {E.~S.}\ \bibnamefont
  {S\o{}rensen}},\ }\bibfield  {title} {\bibinfo {title} {Entanglement entropy
  in quantum impurity systems and systems with boundaries},\ }\href
  {https://doi.org/10.1088/1751-8113/42/50/504009} {\bibfield  {journal}
  {\bibinfo  {journal} {J. Phys. A: Math. Theor.}\ }\textbf {\bibinfo {volume}
  {42}},\ \bibinfo {pages} {504009} (\bibinfo {year} {2009})}\BibitemShut
  {NoStop}%
\bibitem [{\citenamefont {Affleck}\ and\ \citenamefont {Ludwig}(1991)}]{AL91}%
  \BibitemOpen
  \bibfield  {author} {\bibinfo {author} {\bibfnamefont {I.}~\bibnamefont
  {Affleck}}\ and\ \bibinfo {author} {\bibfnamefont {A.~W.~W.}\ \bibnamefont
  {Ludwig}},\ }\bibfield  {title} {\bibinfo {title} {Universal noninteger
  ``ground-state degeneracy'' in critical quantum systems},\ }\href
  {https://doi.org/10.1103/PhysRevLett.67.161} {\bibfield  {journal} {\bibinfo
  {journal} {Phys. Rev. Lett.}\ }\textbf {\bibinfo {volume} {67}},\ \bibinfo
  {pages} {161} (\bibinfo {year} {1991})}\BibitemShut {NoStop}%
\bibitem [{\citenamefont {Taddia}\ \emph {et~al.}(2013)\citenamefont {Taddia},
  \citenamefont {Xavier}, \citenamefont {Alcaraz},\ and\ \citenamefont
  {Sierra}}]{txas-13}%
  \BibitemOpen
  \bibfield  {author} {\bibinfo {author} {\bibfnamefont {L.}~\bibnamefont
  {Taddia}}, \bibinfo {author} {\bibfnamefont {J.}~\bibnamefont {Xavier}},
  \bibinfo {author} {\bibfnamefont {F.~C.}\ \bibnamefont {Alcaraz}},\ and\
  \bibinfo {author} {\bibfnamefont {G.}~\bibnamefont {Sierra}},\ }\bibfield
  {title} {\bibinfo {title} {Entanglement entropies in conformal systems with
  boundaries},\ }\href {https://doi.org/10.1103/PhysRevB.88.075112} {\bibfield
  {journal} {\bibinfo  {journal} {Phys. Rev. B}\ }\textbf {\bibinfo {volume}
  {88}},\ \bibinfo {pages} {075112} (\bibinfo {year} {2013})}\BibitemShut
  {NoStop}%
\bibitem [{\citenamefont {Estienne}\ \emph {et~al.}(2022)\citenamefont
  {Estienne}, \citenamefont {Ikhlef},\ and\ \citenamefont {Rotaru}}]{eir-22}%
  \BibitemOpen
  \bibfield  {author} {\bibinfo {author} {\bibfnamefont {B.}~\bibnamefont
  {Estienne}}, \bibinfo {author} {\bibfnamefont {Y.}~\bibnamefont {Ikhlef}},\
  and\ \bibinfo {author} {\bibfnamefont {A.}~\bibnamefont {Rotaru}},\
  }\bibfield  {title} {\bibinfo {title} {{Second R\'enyi entropy and annulus
  partition function for one-dimensional quantum critical systems with
  boundaries}},\ }\href {https://doi.org/10.21468/SciPostPhys.12.4.141}
  {\bibfield  {journal} {\bibinfo  {journal} {SciPost Phys.}\ }\textbf
  {\bibinfo {volume} {12}},\ \bibinfo {pages} {141} (\bibinfo {year}
  {2022})}\BibitemShut {NoStop}%
\bibitem [{\citenamefont {Estienne}\ \emph {et~al.}(2023)\citenamefont
  {Estienne}, \citenamefont {Ikhlef},\ and\ \citenamefont {Rotaru}}]{eir-23}%
  \BibitemOpen
  \bibfield  {author} {\bibinfo {author} {\bibfnamefont {B.}~\bibnamefont
  {Estienne}}, \bibinfo {author} {\bibfnamefont {Y.}~\bibnamefont {Ikhlef}},\
  and\ \bibinfo {author} {\bibfnamefont {A.}~\bibnamefont {Rotaru}},\ }\href
  {https://doi.org/10.48550/arXiv.2301.02124} {\bibinfo {title} {R\'enyi
  entropies for one-dimensional quantum systems with mixed boundary
  conditions}} (\bibinfo {year} {2023}),\ \bibinfo {note}
  {arXiv:2301.02124}\BibitemShut {NoStop}%
\bibitem [{\citenamefont {Taddia}\ \emph {et~al.}(2016)\citenamefont {Taddia},
  \citenamefont {Ortolani},\ and\ \citenamefont {P{\'{a}}lmai}}]{top-16}%
  \BibitemOpen
  \bibfield  {author} {\bibinfo {author} {\bibfnamefont {L.}~\bibnamefont
  {Taddia}}, \bibinfo {author} {\bibfnamefont {F.}~\bibnamefont {Ortolani}},\
  and\ \bibinfo {author} {\bibfnamefont {T.}~\bibnamefont {P{\'{a}}lmai}},\
  }\bibfield  {title} {\bibinfo {title} {R\'enyi entanglement entropies of
  descendant states in critical systems with boundaries: conformal field theory
  and spin chains},\ }\href {https://doi.org/10.1088/1742-5468/2016/09/093104}
  {\bibfield  {journal} {\bibinfo  {journal} {J. Stat. Mech.}\ }\textbf
  {\bibinfo {volume} {2016}},\ \bibinfo {pages} {093104} (\bibinfo {year}
  {2016})}\BibitemShut {NoStop}%
\bibitem [{\citenamefont {Mintchev}\ and\ \citenamefont {Tonni}(2021)}]{mt-20}%
  \BibitemOpen
  \bibfield  {author} {\bibinfo {author} {\bibfnamefont {M.}~\bibnamefont
  {Mintchev}}\ and\ \bibinfo {author} {\bibfnamefont {E.}~\bibnamefont
  {Tonni}},\ }\bibfield  {title} {\bibinfo {title} {{Modular Hamiltonians for
  the massless Dirac field in the presence of a boundary}},\ }\href
  {https://doi.org/10.1007/jhep03(2021)204} {\bibfield  {journal} {\bibinfo
  {journal} {J. High Energ. Phys.}\ }\textbf {\bibinfo {volume} {2021}}\bibinfo
   {number} { (3)},\ \bibinfo {pages} {204}}\BibitemShut {NoStop}%
\bibitem [{\citenamefont {Laflorencie}\ \emph {et~al.}(2006)\citenamefont
  {Laflorencie}, \citenamefont {S\o{}rensen}, \citenamefont {Chang},\ and\
  \citenamefont {Affleck}}]{LSCA06}%
  \BibitemOpen
\bibfield  {number} {  }\bibfield  {author} {\bibinfo {author} {\bibfnamefont
  {N.}~\bibnamefont {Laflorencie}}, \bibinfo {author} {\bibfnamefont {E.~S.}\
  \bibnamefont {S\o{}rensen}}, \bibinfo {author} {\bibfnamefont {M.-S.}\
  \bibnamefont {Chang}},\ and\ \bibinfo {author} {\bibfnamefont
  {I.}~\bibnamefont {Affleck}},\ }\bibfield  {title} {\bibinfo {title}
  {Boundary effects in the critical scaling of entanglement entropy in 1d
  systems},\ }\href {https://doi.org/10.1103/PhysRevLett.96.100603} {\bibfield
  {journal} {\bibinfo  {journal} {Phys. Rev. Lett.}\ }\textbf {\bibinfo
  {volume} {96}},\ \bibinfo {pages} {100603} (\bibinfo {year}
  {2006})}\BibitemShut {NoStop}%
\bibitem [{\citenamefont {Xavier}\ and\ \citenamefont
  {Rajabpour}(2020)}]{XR20}%
  \BibitemOpen
  \bibfield  {author} {\bibinfo {author} {\bibfnamefont {J.~C.}\ \bibnamefont
  {Xavier}}\ and\ \bibinfo {author} {\bibfnamefont {M.~A.}\ \bibnamefont
  {Rajabpour}},\ }\bibfield  {title} {\bibinfo {title} {Entanglement and
  boundary entropy in quantum spin chains with arbitrary direction of the
  boundary magnetic fields},\ }\href
  {https://doi.org/10.1103/PhysRevB.101.235127} {\bibfield  {journal} {\bibinfo
   {journal} {Phys. Rev. B}\ }\textbf {\bibinfo {volume} {101}},\ \bibinfo
  {pages} {235127} (\bibinfo {year} {2020})}\BibitemShut {NoStop}%
\bibitem [{\citenamefont {Roy}\ and\ \citenamefont
  {Saleur}(2022{\natexlab{a}})}]{RS22b}%
  \BibitemOpen
  \bibfield  {author} {\bibinfo {author} {\bibfnamefont {A.}~\bibnamefont
  {Roy}}\ and\ \bibinfo {author} {\bibfnamefont {H.}~\bibnamefont {Saleur}},\
  }\bibinfo {title} {Entanglement entropy in critical quantum spin chains with
  boundaries and defects},\ in\ \href
  {https://doi.org/10.1007/978-3-031-03998-0_3} {\emph {\bibinfo {booktitle}
  {Entanglement in Spin Chains: From Theory to Quantum Technology
  Applications}}},\ \bibinfo {editor} {edited by\ \bibinfo {editor}
  {\bibfnamefont {A.}~\bibnamefont {Bayat}}, \bibinfo {editor} {\bibfnamefont
  {S.}~\bibnamefont {Bose}},\ and\ \bibinfo {editor} {\bibfnamefont
  {H.}~\bibnamefont {Johannesson}}}\ (\bibinfo  {publisher} {Springer
  International Publishing},\ \bibinfo {address} {Cham},\ \bibinfo {year}
  {2022})\ pp.\ \bibinfo {pages} {41--60}\BibitemShut {NoStop}%
\bibitem [{\citenamefont {Herzog}\ and\ \citenamefont
  {Nishioka}(2013)}]{hn-13}%
  \BibitemOpen
  \bibfield  {author} {\bibinfo {author} {\bibfnamefont {C.~P.}\ \bibnamefont
  {Herzog}}\ and\ \bibinfo {author} {\bibfnamefont {T.}~\bibnamefont
  {Nishioka}},\ }\bibfield  {title} {\bibinfo {title} {Entanglement entropy of
  a massive fermion on a torus},\ }\href
  {https://doi.org/10.1007/jhep03(2013)077} {\bibfield  {journal} {\bibinfo
  {journal} {J. High Energ. Phys.}\ }\textbf {\bibinfo {volume} {2013}}\bibinfo
   {number} { (3)},\ \bibinfo {pages} {77}}\BibitemShut {NoStop}%
\bibitem [{\citenamefont {Klich}\ \emph {et~al.}(2017)\citenamefont {Klich},
  \citenamefont {Vaman},\ and\ \citenamefont {Wong}}]{KVW17}%
  \BibitemOpen
\bibfield  {number} {  }\bibfield  {author} {\bibinfo {author} {\bibfnamefont
  {I.}~\bibnamefont {Klich}}, \bibinfo {author} {\bibfnamefont
  {D.}~\bibnamefont {Vaman}},\ and\ \bibinfo {author} {\bibfnamefont
  {G.}~\bibnamefont {Wong}},\ }\bibfield  {title} {\bibinfo {title}
  {Entanglement hamiltonians for chiral fermions with zero modes},\ }\href
  {https://doi.org/10.1103/PhysRevLett.119.120401} {\bibfield  {journal}
  {\bibinfo  {journal} {Phys. Rev. Lett.}\ }\textbf {\bibinfo {volume} {119}},\
  \bibinfo {pages} {120401} (\bibinfo {year} {2017})}\BibitemShut {NoStop}%
\bibitem [{\citenamefont {Klich}\ \emph {et~al.}(2018)\citenamefont {Klich},
  \citenamefont {Vaman},\ and\ \citenamefont {Wong}}]{KVW18}%
  \BibitemOpen
  \bibfield  {author} {\bibinfo {author} {\bibfnamefont {I.}~\bibnamefont
  {Klich}}, \bibinfo {author} {\bibfnamefont {D.}~\bibnamefont {Vaman}},\ and\
  \bibinfo {author} {\bibfnamefont {G.}~\bibnamefont {Wong}},\ }\bibfield
  {title} {\bibinfo {title} {Entanglement hamiltonians and entropy in 1 + 1d
  chiral fermion systems},\ }\href {https://doi.org/10.1103/PhysRevB.98.035134}
  {\bibfield  {journal} {\bibinfo  {journal} {Phys. Rev. B}\ }\textbf {\bibinfo
  {volume} {98}},\ \bibinfo {pages} {035134} (\bibinfo {year}
  {2018})}\BibitemShut {NoStop}%
\bibitem [{\citenamefont {Roy}\ and\ \citenamefont
  {Saleur}(2022{\natexlab{b}})}]{RS22}%
  \BibitemOpen
  \bibfield  {author} {\bibinfo {author} {\bibfnamefont {A.}~\bibnamefont
  {Roy}}\ and\ \bibinfo {author} {\bibfnamefont {H.}~\bibnamefont {Saleur}},\
  }\bibfield  {title} {\bibinfo {title} {{Entanglement Entropy in the Ising
  Model with Topological Defects}},\ }\href
  {https://doi.org/10.1103/PhysRevLett.128.090603} {\bibfield  {journal}
  {\bibinfo  {journal} {Phys. Rev. Lett.}\ }\textbf {\bibinfo {volume} {128}},\
  \bibinfo {pages} {090603} (\bibinfo {year} {2022}{\natexlab{b}})}\BibitemShut
  {NoStop}%
\bibitem [{\citenamefont {Rogerson}\ \emph {et~al.}(2022)\citenamefont
  {Rogerson}, \citenamefont {Pollmann},\ and\ \citenamefont {Roy}}]{rpr-22}%
  \BibitemOpen
  \bibfield  {author} {\bibinfo {author} {\bibfnamefont {D.}~\bibnamefont
  {Rogerson}}, \bibinfo {author} {\bibfnamefont {F.}~\bibnamefont {Pollmann}},\
  and\ \bibinfo {author} {\bibfnamefont {A.}~\bibnamefont {Roy}},\ }\bibfield
  {title} {\bibinfo {title} {{Entanglement entropy and negativity in the Ising
  model with defects}},\ }\href {https://doi.org/10.1007/jhep06(2022)165}
  {\bibfield  {journal} {\bibinfo  {journal} {J. High Energ. Phys.}\ }\textbf
  {\bibinfo {volume} {2022}}\bibinfo  {number} { (6)},\ \bibinfo {pages}
  {165}}\BibitemShut {NoStop}%
\bibitem [{\citenamefont {Aasen}\ \emph {et~al.}(2016)\citenamefont {Aasen},
  \citenamefont {Mong},\ and\ \citenamefont {Fendley}}]{amf-16}%
  \BibitemOpen
\bibfield  {number} {  }\bibfield  {author} {\bibinfo {author} {\bibfnamefont
  {D.}~\bibnamefont {Aasen}}, \bibinfo {author} {\bibfnamefont {R.~S.~K.}\
  \bibnamefont {Mong}},\ and\ \bibinfo {author} {\bibfnamefont
  {P.}~\bibnamefont {Fendley}},\ }\bibfield  {title} {\bibinfo {title}
  {{Topological defects on the lattice: I. The Ising model}},\ }\href
  {https://doi.org/10.1088/1751-8113/49/35/354001} {\bibfield  {journal}
  {\bibinfo  {journal} {J. Phys. A: Math. Theor.}\ }\textbf {\bibinfo {volume}
  {49}},\ \bibinfo {pages} {354001} (\bibinfo {year} {2016})}\BibitemShut
  {NoStop}%
\bibitem [{\citenamefont {Kitaev}(2001)}]{Kitaev01}%
  \BibitemOpen
  \bibfield  {author} {\bibinfo {author} {\bibfnamefont {A.~Y.}\ \bibnamefont
  {Kitaev}},\ }\bibfield  {title} {\bibinfo {title} {Unpaired majorana fermions
  in quantum wires},\ }\href {https://doi.org/10.1070/1063-7869/44/10S/S29}
  {\bibfield  {journal} {\bibinfo  {journal} {Physics-Uspekhi}\ }\textbf
  {\bibinfo {volume} {44}},\ \bibinfo {pages} {131} (\bibinfo {year}
  {2001})}\BibitemShut {NoStop}%
\bibitem [{\citenamefont {Fendley}(2012)}]{Fendley12}%
  \BibitemOpen
  \bibfield  {author} {\bibinfo {author} {\bibfnamefont {P.}~\bibnamefont
  {Fendley}},\ }\bibfield  {title} {\bibinfo {title} {{Parafermionic edge zero
  modes in Zn-invariant spin chains}},\ }\href
  {https://doi.org/10.1088/1742-5468/2012/11/P11020} {\bibfield  {journal}
  {\bibinfo  {journal} {J. Stat. Mech.: Theory Exp.}\ }\textbf {\bibinfo
  {volume} {2012}}\bibinfo  {number} { (11)},\ \bibinfo {pages}
  {P11020}}\BibitemShut {NoStop}%
\bibitem [{\citenamefont {Fendley}(2016)}]{Fendley16}%
  \BibitemOpen
\bibfield  {number} {  }\bibfield  {author} {\bibinfo {author} {\bibfnamefont
  {P.}~\bibnamefont {Fendley}},\ }\bibfield  {title} {\bibinfo {title} {Strong
  zero modes and eigenstate phase transitions in the {XYZ}/interacting
  {Majorana} chain},\ }\href {https://doi.org/10.1088/1751-8113/49/30/30LT01}
  {\bibfield  {journal} {\bibinfo  {journal} {J. Phys. A: Math. Theor.}\
  }\textbf {\bibinfo {volume} {49}},\ \bibinfo {pages} {30LT01} (\bibinfo
  {year} {2016})}\BibitemShut {NoStop}%
\bibitem [{\citenamefont {Peschel}(2005)}]{peschel-05}%
  \BibitemOpen
  \bibfield  {author} {\bibinfo {author} {\bibfnamefont {I.}~\bibnamefont
  {Peschel}},\ }\bibfield  {title} {\bibinfo {title} {Entanglement entropy with
  interface defects},\ }\href {https://doi.org/10.1088/0305-4470/38/20/002}
  {\bibfield  {journal} {\bibinfo  {journal} {J. Phys. A: Math. Gen.}\ }\textbf
  {\bibinfo {volume} {38}},\ \bibinfo {pages} {4327} (\bibinfo {year}
  {2005})}\BibitemShut {NoStop}%
\bibitem [{\citenamefont {Igl\'oi}\ \emph {et~al.}(2009)\citenamefont
  {Igl\'oi}, \citenamefont {Szatm\'ari},\ and\ \citenamefont {Lin}}]{isl-09}%
  \BibitemOpen
  \bibfield  {author} {\bibinfo {author} {\bibfnamefont {F.}~\bibnamefont
  {Igl\'oi}}, \bibinfo {author} {\bibfnamefont {Z.}~\bibnamefont
  {Szatm\'ari}},\ and\ \bibinfo {author} {\bibfnamefont {Y.-C.}\ \bibnamefont
  {Lin}},\ }\bibfield  {title} {\bibinfo {title} {Entanglement entropy with
  localized and extended interface defects},\ }\href
  {https://doi.org/10.1103/PhysRevB.80.024405} {\bibfield  {journal} {\bibinfo
  {journal} {Phys. Rev. B}\ }\textbf {\bibinfo {volume} {80}},\ \bibinfo
  {pages} {024405} (\bibinfo {year} {2009})}\BibitemShut {NoStop}%
\bibitem [{\citenamefont {Eisler}\ and\ \citenamefont {Garmon}(2010)}]{eg-10}%
  \BibitemOpen
  \bibfield  {author} {\bibinfo {author} {\bibfnamefont {V.}~\bibnamefont
  {Eisler}}\ and\ \bibinfo {author} {\bibfnamefont {S.~S.}\ \bibnamefont
  {Garmon}},\ }\bibfield  {title} {\bibinfo {title} {Fano resonances and
  entanglement entropy},\ }\href {https://doi.org/10.1103/PhysRevB.82.174202}
  {\bibfield  {journal} {\bibinfo  {journal} {Phys. Rev. B}\ }\textbf {\bibinfo
  {volume} {82}},\ \bibinfo {pages} {174202} (\bibinfo {year}
  {2010})}\BibitemShut {NoStop}%
\bibitem [{\citenamefont {Eisler}\ and\ \citenamefont {Peschel}(2010)}]{ep-10}%
  \BibitemOpen
  \bibfield  {author} {\bibinfo {author} {\bibfnamefont {V.}~\bibnamefont
  {Eisler}}\ and\ \bibinfo {author} {\bibfnamefont {I.}~\bibnamefont
  {Peschel}},\ }\bibfield  {title} {\bibinfo {title} {Entanglement in fermionic
  chains with interface defects},\ }\href
  {https://doi.org/10.1002/andp.201000055} {\bibfield  {journal} {\bibinfo
  {journal} {Ann. Phys. (Berlin)}\ }\textbf {\bibinfo {volume} {522}},\
  \bibinfo {pages} {679} (\bibinfo {year} {2010})}\BibitemShut {NoStop}%
\bibitem [{\citenamefont {Peschel}\ and\ \citenamefont {Eisler}(2012)}]{pe-12}%
  \BibitemOpen
  \bibfield  {author} {\bibinfo {author} {\bibfnamefont {I.}~\bibnamefont
  {Peschel}}\ and\ \bibinfo {author} {\bibfnamefont {V.}~\bibnamefont
  {Eisler}},\ }\bibfield  {title} {\bibinfo {title} {Exact results for the
  entanglement across defects in critical chains},\ }\href
  {https://doi.org/10.1088/1751-8113/45/15/155301} {\bibfield  {journal}
  {\bibinfo  {journal} {J. Phys. A: Math. Theor.}\ }\textbf {\bibinfo {volume}
  {45}},\ \bibinfo {pages} {155301} (\bibinfo {year} {2012})}\BibitemShut
  {NoStop}%
\bibitem [{\citenamefont {Sakai}\ and\ \citenamefont {Satoh}(2008)}]{ss-08}%
  \BibitemOpen
  \bibfield  {author} {\bibinfo {author} {\bibfnamefont {K.}~\bibnamefont
  {Sakai}}\ and\ \bibinfo {author} {\bibfnamefont {Y.}~\bibnamefont {Satoh}},\
  }\bibfield  {title} {\bibinfo {title} {Entanglement through conformal
  interfaces},\ }\href {https://doi.org/10.1088/1126-6708/2008/12/001}
  {\bibfield  {journal} {\bibinfo  {journal} {J. High Energy Phys.}\ }\textbf
  {\bibinfo {volume} {12}}\bibinfo  {number} { (12)},\ \bibinfo {pages}
  {001}}\BibitemShut {NoStop}%
\bibitem [{\citenamefont {Brehm}\ and\ \citenamefont {Brunner}(2015)}]{bb-15}%
  \BibitemOpen
\bibfield  {number} {  }\bibfield  {author} {\bibinfo {author} {\bibfnamefont
  {E.}~\bibnamefont {Brehm}}\ and\ \bibinfo {author} {\bibfnamefont
  {I.}~\bibnamefont {Brunner}},\ }\bibfield  {title} {\bibinfo {title}
  {{Entanglement entropy through conformal interfaces in the 2D Ising model}},\
  }\href {https://doi.org/10.1007/JHEP09(2015)080} {\bibfield  {journal}
  {\bibinfo  {journal} {J. High Energ. Phys.}\ }\textbf {\bibinfo {volume}
  {2015}}\bibinfo  {number} { (9)},\ \bibinfo {pages} {80}}\BibitemShut
  {NoStop}%
\bibitem [{\citenamefont {Gutperle}\ and\ \citenamefont
  {Miller}(2017)}]{gm-17}%
  \BibitemOpen
\bibfield  {number} {  }\bibfield  {author} {\bibinfo {author} {\bibfnamefont
  {M.}~\bibnamefont {Gutperle}}\ and\ \bibinfo {author} {\bibfnamefont {J.~D.}\
  \bibnamefont {Miller}},\ }\bibfield  {title} {\bibinfo {title} {Entanglement
  entropy at {CFT} junctions},\ }\href
  {https://doi.org/10.1103/physrevd.95.106008} {\bibfield  {journal} {\bibinfo
  {journal} {Phys. Rev. D}\ }\textbf {\bibinfo {volume} {95}},\ \bibinfo
  {pages} {106008} (\bibinfo {year} {2017})}\BibitemShut {NoStop}%
\bibitem [{\citenamefont {Capizzi}\ \emph
  {et~al.}(2022{\natexlab{a}})\citenamefont {Capizzi}, \citenamefont
  {Murciano},\ and\ \citenamefont {Calabrese}}]{cmc-22}%
  \BibitemOpen
  \bibfield  {author} {\bibinfo {author} {\bibfnamefont {L.}~\bibnamefont
  {Capizzi}}, \bibinfo {author} {\bibfnamefont {S.}~\bibnamefont {Murciano}},\
  and\ \bibinfo {author} {\bibfnamefont {P.}~\bibnamefont {Calabrese}},\
  }\bibfield  {title} {\bibinfo {title} {Rényi entropy and negativity for
  massless dirac fermions at conformal interfaces and junctions},\ }\href
  {https://doi.org/10.1007/JHEP08(2022)171} {\bibfield  {journal} {\bibinfo
  {journal} {J. High Energ. Phys.}\ }\textbf {\bibinfo {volume} {2022}}\bibinfo
   {number} { (8)},\ \bibinfo {pages} {171}}\BibitemShut {NoStop}%
\bibitem [{\citenamefont {Capizzi}\ \emph
  {et~al.}(2022{\natexlab{b}})\citenamefont {Capizzi}, \citenamefont
  {Murciano},\ and\ \citenamefont {Calabrese}}]{cmc-22a}%
  \BibitemOpen
\bibfield  {number} {  }\bibfield  {author} {\bibinfo {author} {\bibfnamefont
  {L.}~\bibnamefont {Capizzi}}, \bibinfo {author} {\bibfnamefont
  {S.}~\bibnamefont {Murciano}},\ and\ \bibinfo {author} {\bibfnamefont
  {P.}~\bibnamefont {Calabrese}},\ }\bibfield  {title} {\bibinfo {title}
  {R{\'{e}}nyi entropy and negativity for massless complex boson at conformal
  interfaces and junctions},\ }\href {https://doi.org/10.1007/jhep11(2022)105}
  {\bibfield  {journal} {\bibinfo  {journal} {J. High Energ. Phys.}\ }\textbf
  {\bibinfo {volume} {2022}}\bibinfo  {number} { (11)},\ \bibinfo {pages}
  {105}}\BibitemShut {NoStop}%
\bibitem [{\citenamefont {Bachas}\ \emph {et~al.}(2002)\citenamefont {Bachas},
  \citenamefont {Boer}, \citenamefont {Dijkgraaf},\ and\ \citenamefont
  {Ooguri}}]{bbdo-02}%
  \BibitemOpen
\bibfield  {number} {  }\bibfield  {author} {\bibinfo {author} {\bibfnamefont
  {C.}~\bibnamefont {Bachas}}, \bibinfo {author} {\bibfnamefont
  {J.}~\bibnamefont {Boer}}, \bibinfo {author} {\bibfnamefont {R.}~\bibnamefont
  {Dijkgraaf}},\ and\ \bibinfo {author} {\bibfnamefont {H.}~\bibnamefont
  {Ooguri}},\ }\bibfield  {title} {\bibinfo {title} {Permeable conformal walls
  and holography},\ }\href {https://doi.org/10.1088/1126-6708/2002/06/027}
  {\bibfield  {journal} {\bibinfo  {journal} {J. High Energ. Phys.}\ }\textbf
  {\bibinfo {volume} {2002}}\bibinfo  {number} { (06)},\ \bibinfo {pages}
  {027}}\BibitemShut {NoStop}%
\bibitem [{\citenamefont {Eisler}\ and\ \citenamefont {Peschel}(2012)}]{EP12}%
  \BibitemOpen
\bibfield  {number} {  }\bibfield  {author} {\bibinfo {author} {\bibfnamefont
  {V.}~\bibnamefont {Eisler}}\ and\ \bibinfo {author} {\bibfnamefont
  {I.}~\bibnamefont {Peschel}},\ }\bibfield  {title} {\bibinfo {title} {On
  entanglement evolution across defects in critical chains},\ }\href
  {https://doi.org/10.1209/0295-5075/99/20001} {\bibfield  {journal} {\bibinfo
  {journal} {Europhys Lett.}\ }\textbf {\bibinfo {volume} {99}},\ \bibinfo
  {pages} {20001} (\bibinfo {year} {2012})}\BibitemShut {NoStop}%
\bibitem [{\citenamefont {Peschel}(2003)}]{Peschel03}%
  \BibitemOpen
  \bibfield  {author} {\bibinfo {author} {\bibfnamefont {I.}~\bibnamefont
  {Peschel}},\ }\bibfield  {title} {\bibinfo {title} {Calculation of reduced
  density matrices from correlation functions},\ }\href
  {https://doi.org/10.1088/0305-4470/36/14/101} {\bibfield  {journal} {\bibinfo
   {journal} {J. Phys. A: Math. Gen.}\ }\textbf {\bibinfo {volume} {36}},\
  \bibinfo {pages} {L205} (\bibinfo {year} {2003})}\BibitemShut {NoStop}%
\bibitem [{\citenamefont {Peschel}\ and\ \citenamefont {Eisler}(2009)}]{PE09}%
  \BibitemOpen
  \bibfield  {author} {\bibinfo {author} {\bibfnamefont {I.}~\bibnamefont
  {Peschel}}\ and\ \bibinfo {author} {\bibfnamefont {V.}~\bibnamefont
  {Eisler}},\ }\bibfield  {title} {\bibinfo {title} {Reduced density matrices
  and entanglement entropy in free lattice models},\ }\href
  {https://doi.org/10.1088/1751-8113/42/50/504003} {\bibfield  {journal}
  {\bibinfo  {journal} {J. Phys. A: Math. Theor.}\ }\textbf {\bibinfo {volume}
  {42}},\ \bibinfo {pages} {504003} (\bibinfo {year} {2009})}\BibitemShut
  {NoStop}%
\bibitem [{\citenamefont {Klich}\ and\ \citenamefont {Levitov}(2009)}]{kl-09}%
  \BibitemOpen
  \bibfield  {author} {\bibinfo {author} {\bibfnamefont {I.}~\bibnamefont
  {Klich}}\ and\ \bibinfo {author} {\bibfnamefont {L.}~\bibnamefont
  {Levitov}},\ }\bibfield  {title} {\bibinfo {title} {Quantum noise as an
  entanglement meter},\ }\href {https://doi.org/10.1103/physrevlett.102.100502}
  {\bibfield  {journal} {\bibinfo  {journal} {Phys. Rev. Lett.}\ }\textbf
  {\bibinfo {volume} {102}},\ \bibinfo {pages} {100502} (\bibinfo {year}
  {2009})}\BibitemShut {NoStop}%
\bibitem [{\citenamefont {Song}\ \emph {et~al.}(2011)\citenamefont {Song},
  \citenamefont {Flindt}, \citenamefont {Rachel}, \citenamefont {Klich},\ and\
  \citenamefont {Le~Hur}}]{Songetal11}%
  \BibitemOpen
  \bibfield  {author} {\bibinfo {author} {\bibfnamefont {H.~F.}\ \bibnamefont
  {Song}}, \bibinfo {author} {\bibfnamefont {C.}~\bibnamefont {Flindt}},
  \bibinfo {author} {\bibfnamefont {S.}~\bibnamefont {Rachel}}, \bibinfo
  {author} {\bibfnamefont {I.}~\bibnamefont {Klich}},\ and\ \bibinfo {author}
  {\bibfnamefont {K.}~\bibnamefont {Le~Hur}},\ }\bibfield  {title} {\bibinfo
  {title} {Entanglement entropy from charge statistics: Exact relations for
  noninteracting many-body systems},\ }\href
  {https://doi.org/10.1103/PhysRevB.83.161408} {\bibfield  {journal} {\bibinfo
  {journal} {Phys. Rev. B}\ }\textbf {\bibinfo {volume} {83}},\ \bibinfo
  {pages} {161408} (\bibinfo {year} {2011})}\BibitemShut {NoStop}%
\bibitem [{\citenamefont {Abanov}\ \emph {et~al.}(2011)\citenamefont {Abanov},
  \citenamefont {Ivanov},\ and\ \citenamefont {Qian}}]{AIQ11}%
  \BibitemOpen
  \bibfield  {author} {\bibinfo {author} {\bibfnamefont {A.~G.}\ \bibnamefont
  {Abanov}}, \bibinfo {author} {\bibfnamefont {D.~A.}\ \bibnamefont {Ivanov}},\
  and\ \bibinfo {author} {\bibfnamefont {Y.}~\bibnamefont {Qian}},\ }\bibfield
  {title} {\bibinfo {title} {{Quantum fluctuations of one-dimensional free
  fermions and Fisher-Hartwig formula for Toeplitz determinants}},\ }\href
  {https://doi.org/10.1088/1751-8113/44/48/485001} {\bibfield  {journal}
  {\bibinfo  {journal} {J. Phys. A: Math. Theor.}\ }\textbf {\bibinfo {volume}
  {44}},\ \bibinfo {pages} {485001} (\bibinfo {year} {2011})}\BibitemShut
  {NoStop}%
\bibitem [{\citenamefont {Ivanov}\ \emph {et~al.}(2013)\citenamefont {Ivanov},
  \citenamefont {Abanov},\ and\ \citenamefont {Cheianov}}]{IAC13}%
  \BibitemOpen
  \bibfield  {author} {\bibinfo {author} {\bibfnamefont {D.~A.}\ \bibnamefont
  {Ivanov}}, \bibinfo {author} {\bibfnamefont {A.~G.}\ \bibnamefont {Abanov}},\
  and\ \bibinfo {author} {\bibfnamefont {V.~V.}\ \bibnamefont {Cheianov}},\
  }\bibfield  {title} {\bibinfo {title} {{Counting free fermions on a line: a
  Fisher-Hartwig asymptotic expansion for the Toeplitz determinant in the
  double-scaling limit}},\ }\href
  {https://doi.org/10.1088/1751-8113/46/8/085003} {\bibfield  {journal}
  {\bibinfo  {journal} {J. Phys. A: Math. Theor.}\ }\textbf {\bibinfo {volume}
  {46}},\ \bibinfo {pages} {085003} (\bibinfo {year} {2013})}\BibitemShut
  {NoStop}%
\bibitem [{\citenamefont {S\"usstrunk}\ and\ \citenamefont
  {Ivanov}(2013)}]{SI13}%
  \BibitemOpen
  \bibfield  {author} {\bibinfo {author} {\bibfnamefont {R.}~\bibnamefont
  {S\"usstrunk}}\ and\ \bibinfo {author} {\bibfnamefont {D.~A.}\ \bibnamefont
  {Ivanov}},\ }\bibfield  {title} {\bibinfo {title} {Free fermions on a line:
  Asymptotics of the entanglement entropy and entanglement spectrum from full
  counting statistics},\ }\href {https://doi.org/10.1209/0295-5075/100/60009}
  {\bibfield  {journal} {\bibinfo  {journal} {Europhys. Lett.}\ }\textbf
  {\bibinfo {volume} {100}},\ \bibinfo {pages} {60009} (\bibinfo {year}
  {2013})}\BibitemShut {NoStop}%
\bibitem [{\citenamefont {Capizzi}\ \emph
  {et~al.}(2023{\natexlab{a}})\citenamefont {Capizzi}, \citenamefont
  {Murciano},\ and\ \citenamefont {Calabrese}}]{cmc-23}%
  \BibitemOpen
  \bibfield  {author} {\bibinfo {author} {\bibfnamefont {L.}~\bibnamefont
  {Capizzi}}, \bibinfo {author} {\bibfnamefont {S.}~\bibnamefont {Murciano}},\
  and\ \bibinfo {author} {\bibfnamefont {P.}~\bibnamefont {Calabrese}},\ }\href
  {https://doi.org/https://doi.org/10.48550/arXiv.2302.08209} {\bibinfo {title}
  {Full counting statistics and symmetry resolved entanglement for free
  conformal theories with interface defects}} (\bibinfo {year}
  {2023}{\natexlab{a}}),\ \bibinfo {note} {arXiv:2302.08209}\BibitemShut
  {NoStop}%
\bibitem [{\citenamefont {Basor}\ and\ \citenamefont {Tracy}(1991)}]{BT91}%
  \BibitemOpen
  \bibfield  {author} {\bibinfo {author} {\bibfnamefont {E.}~\bibnamefont
  {Basor}}\ and\ \bibinfo {author} {\bibfnamefont {C.~A.}\ \bibnamefont
  {Tracy}},\ }\bibfield  {title} {\bibinfo {title} {{The Fisher-Hartwig
  conjecture and generalizations}},\ }\href
  {https://doi.org/10.1016/0378-4371(91)90149-7} {\bibfield  {journal}
  {\bibinfo  {journal} {Physica A}\ }\textbf {\bibinfo {volume} {177}},\
  \bibinfo {pages} {167} (\bibinfo {year} {1991})}\BibitemShut {NoStop}%
\bibitem [{\citenamefont {Deift}\ \emph {et~al.}(2011)\citenamefont {Deift},
  \citenamefont {Its},\ and\ \citenamefont {Krasovsky}}]{DIK11}%
  \BibitemOpen
  \bibfield  {author} {\bibinfo {author} {\bibfnamefont {P.}~\bibnamefont
  {Deift}}, \bibinfo {author} {\bibfnamefont {A.}~\bibnamefont {Its}},\ and\
  \bibinfo {author} {\bibfnamefont {I.}~\bibnamefont {Krasovsky}},\ }\bibfield
  {title} {\bibinfo {title} {{Asymptotics of Toeplitz, Hankel, and
  Toeplitz+Hankel determinants with Fisher-Hartwig singularities}},\ }\href
  {https://doi.org/10.4007/annals.2011.174.2.12} {\bibfield  {journal}
  {\bibinfo  {journal} {Ann. Math.}\ }\textbf {\bibinfo {volume} {174}},\
  \bibinfo {pages} {1243} (\bibinfo {year} {2011})}\BibitemShut {NoStop}%
\bibitem [{\citenamefont {Casini}\ \emph {et~al.}(2005)\citenamefont {Casini},
  \citenamefont {Fosco},\ and\ \citenamefont {Huerta}}]{CFH05}%
  \BibitemOpen
  \bibfield  {author} {\bibinfo {author} {\bibfnamefont {H.}~\bibnamefont
  {Casini}}, \bibinfo {author} {\bibfnamefont {C.~D.}\ \bibnamefont {Fosco}},\
  and\ \bibinfo {author} {\bibfnamefont {M.}~\bibnamefont {Huerta}},\
  }\bibfield  {title} {\bibinfo {title} {Entanglement and alpha entropies for a
  massive {Dirac} field in two dimensions},\ }\href
  {https://doi.org/10.1088/1742-5468/2005/07/P07007} {\bibfield  {journal}
  {\bibinfo  {journal} {J. Stat. Mech.: Theory Exp.}\ }\textbf {\bibinfo
  {volume} {2005}}\bibinfo  {number} { (07)},\ \bibinfo {pages}
  {P07007}}\BibitemShut {NoStop}%
\bibitem [{\citenamefont {Fagotti}\ and\ \citenamefont
  {Calabrese}(2011)}]{FC11}%
  \BibitemOpen
\bibfield  {number} {  }\bibfield  {author} {\bibinfo {author} {\bibfnamefont
  {M.}~\bibnamefont {Fagotti}}\ and\ \bibinfo {author} {\bibfnamefont
  {P.}~\bibnamefont {Calabrese}},\ }\bibfield  {title} {\bibinfo {title}
  {Universal parity effects in the entanglement entropy of {XX} chains with
  open boundary conditions},\ }\href
  {https://doi.org/10.1088/1742-5468/2011/01/p01017} {\bibfield  {journal}
  {\bibinfo  {journal} {J. Stat. Mech.}\ }\textbf {\bibinfo {volume} {2011}},\
  \bibinfo {eid} {P01017} (\bibinfo {year} {2011})}\BibitemShut {NoStop}%
\bibitem [{\citenamefont {Eisler}\ and\ \citenamefont {Peschel}(2013)}]{EP13}%
  \BibitemOpen
  \bibfield  {author} {\bibinfo {author} {\bibfnamefont {V.}~\bibnamefont
  {Eisler}}\ and\ \bibinfo {author} {\bibfnamefont {I.}~\bibnamefont
  {Peschel}},\ }\bibfield  {title} {\bibinfo {title} {Free-fermion entanglement
  and spheroidal functions},\ }\href
  {https://doi.org/10.1088/1742-5468/2013/04/p04028} {\bibfield  {journal}
  {\bibinfo  {journal} {J. Stat. Mech.}\ }\textbf {\bibinfo {volume} {2013}},\
  \bibinfo {eid} {P04028} (\bibinfo {year} {2013})}\BibitemShut {NoStop}%
\bibitem [{\citenamefont {Slepian}(1978)}]{Slepian78}%
  \BibitemOpen
  \bibfield  {author} {\bibinfo {author} {\bibfnamefont {D.}~\bibnamefont
  {Slepian}},\ }\bibfield  {title} {\bibinfo {title} {{Prolate spheroidal wave
  functions, Fourier analysis, and uncertainty - {V: the discrete case}}},\
  }\href {https://doi.org/10.1002/j.1538-7305.1978.tb02104.x} {\bibfield
  {journal} {\bibinfo  {journal} {Bell Syst. Techn. J.}\ }\textbf {\bibinfo
  {volume} {57}},\ \bibinfo {pages} {1371} (\bibinfo {year}
  {1978})}\BibitemShut {NoStop}%
\bibitem [{\citenamefont {Jin}\ and\ \citenamefont {Korepin}(2004)}]{JK04}%
  \BibitemOpen
  \bibfield  {author} {\bibinfo {author} {\bibfnamefont {B.~Q.}\ \bibnamefont
  {Jin}}\ and\ \bibinfo {author} {\bibfnamefont {V.~E.}\ \bibnamefont
  {Korepin}},\ }\bibfield  {title} {\bibinfo {title} {{Quantum spin chain,
  Toeplitz determinants and the Fisher-Hartwig conjecture}},\ }\href
  {https://doi.org/10.1023/B:JOSS.0000037230.37166.42} {\bibfield  {journal}
  {\bibinfo  {journal} {J. Stat. Phys.}\ }\textbf {\bibinfo {volume} {116}},\
  \bibinfo {pages} {79} (\bibinfo {year} {2004})}\BibitemShut {NoStop}%
\bibitem [{\citenamefont {Schlömer}\ \emph {et~al.}(2022)\citenamefont
  {Schlömer}, \citenamefont {Tan}, \citenamefont {Haas},\ and\ \citenamefont
  {Saleur}}]{STHS22}%
  \BibitemOpen
  \bibfield  {author} {\bibinfo {author} {\bibfnamefont {H.}~\bibnamefont
  {Schlömer}}, \bibinfo {author} {\bibfnamefont {C.}~\bibnamefont {Tan}},
  \bibinfo {author} {\bibfnamefont {S.}~\bibnamefont {Haas}},\ and\ \bibinfo
  {author} {\bibfnamefont {H.}~\bibnamefont {Saleur}},\ }\bibfield  {title}
  {\bibinfo {title} {{Parity effects and universal terms of $\mathcal{O}(1)$ in
  the entanglement near a boundary}},\ }\href
  {https://doi.org/10.21468/SciPostPhys.13.5.110} {\bibfield  {journal}
  {\bibinfo  {journal} {SciPost Phys.}\ }\textbf {\bibinfo {volume} {13}},\
  \bibinfo {pages} {110} (\bibinfo {year} {2022})}\BibitemShut {NoStop}%
\bibitem [{\citenamefont {Calabrese}\ \emph {et~al.}(2012)\citenamefont
  {Calabrese}, \citenamefont {Mintchev},\ and\ \citenamefont {Vicari}}]{CMV12}%
  \BibitemOpen
  \bibfield  {author} {\bibinfo {author} {\bibfnamefont {P.}~\bibnamefont
  {Calabrese}}, \bibinfo {author} {\bibfnamefont {M.}~\bibnamefont
  {Mintchev}},\ and\ \bibinfo {author} {\bibfnamefont {E.}~\bibnamefont
  {Vicari}},\ }\bibfield  {title} {\bibinfo {title} {Entanglement entropy of
  quantum wire junctions},\ }\href
  {https://doi.org/10.1088/1751-8113/45/10/105206} {\bibfield  {journal}
  {\bibinfo  {journal} {J. Phys. A: Math. Theor.}\ }\textbf {\bibinfo {volume}
  {45}},\ \bibinfo {pages} {105206} (\bibinfo {year} {2012})}\BibitemShut
  {NoStop}%
\bibitem [{\citenamefont {Gouraud}\ \emph
  {et~al.}(2022{\natexlab{a}})\citenamefont {Gouraud}, \citenamefont
  {Doussal},\ and\ \citenamefont {Schehr}}]{gds-22}%
  \BibitemOpen
  \bibfield  {author} {\bibinfo {author} {\bibfnamefont {G.}~\bibnamefont
  {Gouraud}}, \bibinfo {author} {\bibfnamefont {P.}~\bibnamefont {Doussal}},\
  and\ \bibinfo {author} {\bibfnamefont {G.}~\bibnamefont {Schehr}},\
  }\bibfield  {title} {\bibinfo {title} {Quench dynamics of noninteracting
  fermions with a delta impurity},\ }\href
  {https://doi.org/10.1088/1751-8121/ac83fb} {\bibfield  {journal} {\bibinfo
  {journal} {J. Phys. A Math. Theor.}\ }\textbf {\bibinfo {volume} {55}},\
  \bibinfo {pages} {395001} (\bibinfo {year} {2022}{\natexlab{a}})}\BibitemShut
  {NoStop}%
\bibitem [{\citenamefont {Gouraud}\ \emph
  {et~al.}(2022{\natexlab{b}})\citenamefont {Gouraud}, \citenamefont
  {Doussal},\ and\ \citenamefont {Schehr}}]{gds-23}%
  \BibitemOpen
  \bibfield  {author} {\bibinfo {author} {\bibfnamefont {G.}~\bibnamefont
  {Gouraud}}, \bibinfo {author} {\bibfnamefont {P.~L.}\ \bibnamefont
  {Doussal}},\ and\ \bibinfo {author} {\bibfnamefont {G.}~\bibnamefont
  {Schehr}},\ }\href {https://doi.org/10.48550/ARXIV.2211.15447} {\bibinfo
  {title} {Stationary time correlations for fermions after a quench in the
  presence of an impurity}} (\bibinfo {year} {2022}{\natexlab{b}}),\ \bibinfo
  {note} {arXiv:2211.15447}\BibitemShut {NoStop}%
\bibitem [{\citenamefont {Gamayun}\ \emph {et~al.}(2020)\citenamefont
  {Gamayun}, \citenamefont {Lychkovskiy},\ and\ \citenamefont {Caux}}]{glc-20}%
  \BibitemOpen
  \bibfield  {author} {\bibinfo {author} {\bibfnamefont {O.}~\bibnamefont
  {Gamayun}}, \bibinfo {author} {\bibfnamefont {O.}~\bibnamefont
  {Lychkovskiy}},\ and\ \bibinfo {author} {\bibfnamefont {J.-S.}\ \bibnamefont
  {Caux}},\ }\bibfield  {title} {\bibinfo {title} {{Fredholm determinants, full
  counting statistics and Loschmidt echo for domain wall profiles in
  one-dimensional free fermionic chains}},\ }\href
  {https://doi.org/10.21468/SciPostPhys.8.3.036} {\bibfield  {journal}
  {\bibinfo  {journal} {SciPost Phys.}\ }\textbf {\bibinfo {volume} {8}},\
  \bibinfo {pages} {036} (\bibinfo {year} {2020})}\BibitemShut {NoStop}%
\bibitem [{\citenamefont {Gamayun}\ \emph {et~al.}(2022)\citenamefont
  {Gamayun}, \citenamefont {Zhuravlev},\ and\ \citenamefont {Iorgov}}]{GZI22}%
  \BibitemOpen
  \bibfield  {author} {\bibinfo {author} {\bibfnamefont {O.}~\bibnamefont
  {Gamayun}}, \bibinfo {author} {\bibfnamefont {Y.}~\bibnamefont {Zhuravlev}},\
  and\ \bibinfo {author} {\bibfnamefont {N.}~\bibnamefont {Iorgov}},\ }\href
  {https://doi.org/10.48550/ARXIV.2211.08330} {\bibinfo {title} {On
  {Landauer--Büttiker} formalism from a quantum quench}} (\bibinfo {year}
  {2022}),\ \bibinfo {note} {arXiv:2211.08330}\BibitemShut {NoStop}%
\bibitem [{\citenamefont {Gruber}\ and\ \citenamefont {Eisler}(2020)}]{GE20}%
  \BibitemOpen
  \bibfield  {author} {\bibinfo {author} {\bibfnamefont {M.}~\bibnamefont
  {Gruber}}\ and\ \bibinfo {author} {\bibfnamefont {V.}~\bibnamefont
  {Eisler}},\ }\bibfield  {title} {\bibinfo {title} {Time evolution of
  entanglement negativity across a defect},\ }\href
  {https://doi.org/10.1088/1751-8121/ab831c} {\bibfield  {journal} {\bibinfo
  {journal} {J. Phys. A: Math. Theor.}\ }\textbf {\bibinfo {volume} {53}},\
  \bibinfo {pages} {205301} (\bibinfo {year} {2020})}\BibitemShut {NoStop}%
\bibitem [{\citenamefont {Capizzi}\ and\ \citenamefont {Eisler}(2022)}]{ce-22}%
  \BibitemOpen
  \bibfield  {author} {\bibinfo {author} {\bibfnamefont {L.}~\bibnamefont
  {Capizzi}}\ and\ \bibinfo {author} {\bibfnamefont {V.}~\bibnamefont
  {Eisler}},\ }\href {https://doi.org/10.48550/ARXIV.2209.03297} {\bibinfo
  {title} {Entanglement evolution after a global quench across a conformal
  defect}} (\bibinfo {year} {2022}),\ \bibinfo {note}
  {arxiv:2209.03297}\BibitemShut {NoStop}%
\bibitem [{\citenamefont {Capizzi}\ \emph
  {et~al.}(2023{\natexlab{b}})\citenamefont {Capizzi}, \citenamefont {Scopa},
  \citenamefont {Rottoli},\ and\ \citenamefont {Calabrese}}]{csrc-23}%
  \BibitemOpen
  \bibfield  {author} {\bibinfo {author} {\bibfnamefont {L.}~\bibnamefont
  {Capizzi}}, \bibinfo {author} {\bibfnamefont {S.}~\bibnamefont {Scopa}},
  \bibinfo {author} {\bibfnamefont {F.}~\bibnamefont {Rottoli}},\ and\ \bibinfo
  {author} {\bibfnamefont {P.}~\bibnamefont {Calabrese}},\ }\bibfield  {title}
  {\bibinfo {title} {Domain wall melting across a defect},\ }\href
  {https://doi.org/10.1209/0295-5075/acb50a} {\bibfield  {journal} {\bibinfo
  {journal} {Europhys. Lett.}\ }\textbf {\bibinfo {volume} {141}},\ \bibinfo
  {pages} {31002} (\bibinfo {year} {2023}{\natexlab{b}})}\BibitemShut {NoStop}%
\bibitem [{\citenamefont {Wen}\ \emph {et~al.}(2018)\citenamefont {Wen},
  \citenamefont {Wang},\ and\ \citenamefont {Ryu}}]{wwr-18}%
  \BibitemOpen
  \bibfield  {author} {\bibinfo {author} {\bibfnamefont {X.}~\bibnamefont
  {Wen}}, \bibinfo {author} {\bibfnamefont {Y.}~\bibnamefont {Wang}},\ and\
  \bibinfo {author} {\bibfnamefont {S.}~\bibnamefont {Ryu}},\ }\bibfield
  {title} {\bibinfo {title} {Entanglement evolution across a conformal
  interface},\ }\href {https://doi.org/10.1088/1751-8121/aab561} {\bibfield
  {journal} {\bibinfo  {journal} {J. Phys. A: Math. Theor.}\ }\textbf {\bibinfo
  {volume} {51}},\ \bibinfo {pages} {195004} (\bibinfo {year}
  {2018})}\BibitemShut {NoStop}%
\bibitem [{\citenamefont {Francesco}\ \emph {et~al.}(1997)\citenamefont
  {Francesco}, \citenamefont {Mathieu},\ and\ \citenamefont
  {Sénéchal}}]{DiFrancesco-97}%
  \BibitemOpen
  \bibfield  {author} {\bibinfo {author} {\bibfnamefont {P.}~\bibnamefont
  {Francesco}}, \bibinfo {author} {\bibfnamefont {P.}~\bibnamefont {Mathieu}},\
  and\ \bibinfo {author} {\bibfnamefont {D.}~\bibnamefont {Sénéchal}},\
  }\href {https://doi.org/10.1007/978-1-4612-2256-9} {\emph {\bibinfo {title}
  {Conformal Field Theory}}}\ (\bibinfo  {publisher} {Springer Science \&
  Business Media},\ \bibinfo {year} {1997})\BibitemShut {NoStop}%
\bibitem [{\citenamefont {Gogolin}\ \emph {et~al.}(2004)\citenamefont
  {Gogolin}, \citenamefont {Nersesyan},\ and\ \citenamefont
  {Tsvelik}}]{gnt-04}%
  \BibitemOpen
  \bibfield  {author} {\bibinfo {author} {\bibfnamefont {A.~O.}\ \bibnamefont
  {Gogolin}}, \bibinfo {author} {\bibfnamefont {A.~A.}\ \bibnamefont
  {Nersesyan}},\ and\ \bibinfo {author} {\bibfnamefont {A.~M.}\ \bibnamefont
  {Tsvelik}},\ }\href@noop {} {\emph {\bibinfo {title} {Bosonization and
  strongly correlated systems}}}\ (\bibinfo  {publisher} {Cambridge university
  press},\ \bibinfo {year} {2004})\BibitemShut {NoStop}%
\bibitem [{Note1()}]{Note1}%
  \BibitemOpen
  \bibinfo {note} {The upper half-plane is invariant under $w\rightarrow
  w+\epsilon $ (real translation) and $w\rightarrow e^{\epsilon }w$ (scaling),
  $\epsilon \in \protect \mathbb {R}$}\BibitemShut {NoStop}%
\bibitem [{\citenamefont {Alcaraz}\ \emph {et~al.}(2011)\citenamefont
  {Alcaraz}, \citenamefont {Berganza},\ and\ \citenamefont {Sierra}}]{acb-11}%
  \BibitemOpen
  \bibfield  {author} {\bibinfo {author} {\bibfnamefont {F.~C.}\ \bibnamefont
  {Alcaraz}}, \bibinfo {author} {\bibfnamefont {M.~I.}\ \bibnamefont
  {Berganza}},\ and\ \bibinfo {author} {\bibfnamefont {G.}~\bibnamefont
  {Sierra}},\ }\bibfield  {title} {\bibinfo {title} {Entanglement of low-energy
  excitations in conformal field theory},\ }\href
  {https://doi.org/10.1103/PhysRevLett.106.201601} {\bibfield  {journal}
  {\bibinfo  {journal} {Phys. Rev. Let.}\ }\textbf {\bibinfo {volume} {106}},\
  \bibinfo {pages} {201601} (\bibinfo {year} {2011})}\BibitemShut {NoStop}%
\bibitem [{\citenamefont {Cardy}(2004)}]{Cardy-04}%
  \BibitemOpen
  \bibfield  {author} {\bibinfo {author} {\bibfnamefont {J.}~\bibnamefont
  {Cardy}},\ }\href
  {https://doi.org/https://doi.org/10.48550/arXiv.hep-th/0411189} {\bibinfo
  {title} {Boundary conformal field theory}} (\bibinfo {year} {2004}),\
  \bibinfo {note} {arxiv:0411189}\BibitemShut {NoStop}%
\bibitem [{\citenamefont {Capizzi}\ \emph {et~al.}(2020)\citenamefont
  {Capizzi}, \citenamefont {Ruggiero},\ and\ \citenamefont
  {Calabrese}}]{crc-20}%
  \BibitemOpen
  \bibfield  {author} {\bibinfo {author} {\bibfnamefont {L.}~\bibnamefont
  {Capizzi}}, \bibinfo {author} {\bibfnamefont {P.}~\bibnamefont {Ruggiero}},\
  and\ \bibinfo {author} {\bibfnamefont {P.}~\bibnamefont {Calabrese}},\
  }\bibfield  {title} {\bibinfo {title} {Symmetry resolved entanglement entropy
  of excited states in a {CFT}},\ }\href
  {https://doi.org/10.1088/1742-5468/ab96b6} {\bibfield  {journal} {\bibinfo
  {journal} {J. Stat. Mech.: Theory Exp.}\ }\textbf {\bibinfo {volume}
  {2020}}\bibinfo  {number} { (7)},\ \bibinfo {pages} {073101}}\BibitemShut
  {NoStop}%
\bibitem [{\citenamefont {Fraenkel}\ and\ \citenamefont
  {Goldstein}(2020)}]{fg-20}%
  \BibitemOpen
\bibfield  {number} {  }\bibfield  {author} {\bibinfo {author} {\bibfnamefont
  {S.}~\bibnamefont {Fraenkel}}\ and\ \bibinfo {author} {\bibfnamefont
  {M.}~\bibnamefont {Goldstein}},\ }\bibfield  {title} {\bibinfo {title}
  {Symmetry resolved entanglement: exact results in 1d and beyond},\ }\href
  {https://doi.org/10.1088/1742-5468/ab7753} {\bibfield  {journal} {\bibinfo
  {journal} {J. Stat. Mech.: Theory Exp.}\ }\textbf {\bibinfo {volume}
  {2020}}\bibinfo  {number} { (3)},\ \bibinfo {pages} {033106}}\BibitemShut
  {NoStop}%
\bibitem [{\citenamefont {Calabrese}\ \emph
  {et~al.}(2011{\natexlab{a}})\citenamefont {Calabrese}, \citenamefont
  {Mintchev},\ and\ \citenamefont {Vicari}}]{cmv-11}%
  \BibitemOpen
\bibfield  {number} {  }\bibfield  {author} {\bibinfo {author} {\bibfnamefont
  {P.}~\bibnamefont {Calabrese}}, \bibinfo {author} {\bibfnamefont
  {M.}~\bibnamefont {Mintchev}},\ and\ \bibinfo {author} {\bibfnamefont
  {E.}~\bibnamefont {Vicari}},\ }\bibfield  {title} {\bibinfo {title} {The
  entanglement entropy of one-dimensional systems in continuous and homogeneous
  space},\ }\href {https://doi.org/10.1088/1742-5468/2011/09/p09028} {\bibfield
   {journal} {\bibinfo  {journal} {J. Stat. Mech.: Theory Exp.}\ }\textbf
  {\bibinfo {volume} {2011}}\bibinfo  {number} { (09)},\ \bibinfo {pages}
  {P09028}}\BibitemShut {NoStop}%
\bibitem [{\citenamefont {Calabrese}\ \emph
  {et~al.}(2011{\natexlab{b}})\citenamefont {Calabrese}, \citenamefont
  {Mintchev},\ and\ \citenamefont {Vicari}}]{cmv-11a}%
  \BibitemOpen
\bibfield  {number} {  }\bibfield  {author} {\bibinfo {author} {\bibfnamefont
  {P.}~\bibnamefont {Calabrese}}, \bibinfo {author} {\bibfnamefont
  {M.}~\bibnamefont {Mintchev}},\ and\ \bibinfo {author} {\bibfnamefont
  {E.}~\bibnamefont {Vicari}},\ }\bibfield  {title} {\bibinfo {title}
  {Entanglement entropy of one-dimensional gases},\ }\href
  {https://doi.org/10.1103/PhysRevLett.107.020601} {\bibfield  {journal}
  {\bibinfo  {journal} {Phys. Rev. Lett.}\ }\textbf {\bibinfo {volume} {107}},\
  \bibinfo {pages} {020601} (\bibinfo {year} {2011}{\natexlab{b}})}\BibitemShut
  {NoStop}%
\end{thebibliography}%

\end{document}